\documentclass[longauth]{aa}

\usepackage{graphicx}
\usepackage{natbib}
\usepackage{scalerel}
\usepackage{booktabs}
\usepackage{csquotes}
\usepackage{soul}
\usepackage[table]{xcolor}
\usepackage{fontawesome}
\usepackage{multirow}	

\usepackage{txfonts}
\usepackage[pdfencoding=auto,psdextra]{hyperref}
\hypersetup{
    colorlinks=true,
    linkcolor=blue,
    filecolor=magenta,      
    urlcolor=blue,
    citecolor=blue
}
\makeatletter
\renewcommand*\aa@pageof{, page \thepage{} of \pageref*{LastPage}}
\makeatother

\usepackage[utf8]{inputenc}

\usepackage[switch, modulo]{lineno}
\usepackage{orcidlink}
\usepackage{euclid}
\newcommand{\lcdm}{$\Lambda$CDM\xspace}
\newcommand{\wcdm}{$w$CDM\xspace}
\newcommand{\om}{\Omega_\mathrm{m}}
\newcommand{\ob}{\Omega_\mathrm{b}}
\newcommand{\ol}{\Omega_\Lambda}
\newcommand{\ode}{\Omega_{\rm de}}

\newcommand{\fRz}{\overline{f_\mathrm{R0}}}
\newcommand{\horc}{H_0r_\mathrm{c}}
\newcommand{\de}{\rm de}
\newcommand{\neff}{N_{\rm eff}}
\newcommand{\kmsmpc}{\, \kmsMpc}
\newcommand{\kpcoh}{\, h^{-1}\, \mathrm{kpc}}
\newcommand{\mpcoh}{\, \hMpc}
\newcommand{\gpcoh}{\, h^{-1}\, \mathrm{Gpc}}
\newcommand{\gpcohcube}{\, h^{-3}\, \mathrm{Gpc}^3}
\newcommand{\homopc}{\, h\,\mathrm{Mpc}^{-1}}
\newcommand{\msoh}{h^{-1}\,\ensuremath{{M}_{\odot}}\,}
\newcommand{\energydensity}{u}
\newcommand{\floor}{\, \mathrm{floor}}

\newcommand{\euclid}{\textit{Euclid}\xspace}

\newcommand{\PGadgetThree}{\texttt{p-GADGET3}\xspace}
\newcommand{\ECOSMOG}{\texttt{ECOSMOG}\xspace}
\newcommand{\COLA}{\texttt{MG-COLA}\xspace}
\newcommand{\RAMSES}{\texttt{RAMSES}\xspace}
\newcommand{\FASTPM}{\texttt{FastPM}\xspace}
\newcommand{\AREPO}{\texttt{Arepo}\xspace}
\newcommand{\MGAREPO}{\texttt{MG-Arepo}\xspace}
\newcommand{\CGADGET}{\texttt{c-GADGET}\xspace}
\newcommand{\LGADGET}{\texttt{L-GADGET}\xspace}
\newcommand{\GADGETTWO}{\texttt{GADGET-2}\xspace}
\newcommand{\GADGET}{\texttt{GADGET}\xspace}
\newcommand{\MGGADGET}{\texttt{MG-GADGET}\xspace}
\newcommand{\GEVOLUTIONONETWO}{\texttt{gevolution-1.2}\xspace}
\newcommand{\GIZMO}{\texttt{GIZMO}\xspace}
\newcommand{\KEVOLUTION}{\texttt{$k$-evolution}\xspace}
\newcommand{\COMPLEMENTARY}{\textsc{Complementary}\xspace}
\newcommand{\DEMNUni}{\textsc{DEMNUni}\xspace}
\newcommand{\RAYGAL}{\textsc{Raygal}\xspace}
\newcommand{\ELEPHANT}{\textsc{Elephant}\xspace}
\newcommand{\ColaHiRes}{\textsc{COLA HiRes}\xspace}
\newcommand{\DUSTGRAIN}{\textsc{DUSTGRAIN}\xspace}
\newcommand{\DUSTGRAINPATHFINDER}{\textsc{DUSTGRAIN-PF}\xspace}
\newcommand{\CIDER}{\textsc{CiDER}\xspace}
\newcommand{\DAKAR}{\textsc{DAKAR}\xspace}
\newcommand{\DAKARTWO}{\textsc{DAKAR2}\xspace}
\newcommand{\DAKARONEANDTWO}{\textsc{DAKAR (1\&2)}\xspace}
\newcommand{\ClusteringDE}{\textsc{Clustering DE}\xspace}
\newcommand{\FORGE}{\textsc{FORGE}\xspace}
\newcommand{\BRIDGE}{\textsc{BRIDGE}\xspace}
\newcommand{\PNGUNITsim}{\textsc{PNG-UNIT}\xspace}
\newcommand{\UNITsim}{\textsc{UNIT}\xspace}

\newcommand{\rockstar}{\texttt{Rockstar}\xspace}
\newcommand{\nbodykit}{\texttt{nbodykit}\xspace}
\newcommand{\VIDE}{\texttt{VIDE}\xspace}
\newcommand{\ZOBOV}{\texttt{ZOBOV}\xspace}
\newcommand{\MPGRAFIC}{\texttt{MPGRAFIC}\xspace}
\newcommand{\NGENIC}{\texttt{N-GenIC}\xspace}
\newcommand{\TWOLPTIC}{\texttt{2LPTic}\xspace}
\newcommand{\CAMB}{\texttt{CAMB}\xspace}
\newcommand{\COMET}{\texttt{COMET}\xspace}
\newcommand{\COLOSSUS}{\texttt{COLOSSUS}\xspace}
\newcommand{\numpy}{\texttt{NumPy}\xspace}
\newcommand{\pandas}{\texttt{Pandas}\xspace}
\newcommand{\polars}{\texttt{Polars}\xspace}

\definecolor{darkgreen}{rgb}{0.,0.5,0}

\newcommand\blfootnote[1]{%
  \begingroup
  \renewcommand\thefootnote{}\footnote{#1}%
  \addtocounter{footnote}{-1}%
  \endgroup
}

\newcommand{\comment}[1]{}

\begin{document} 
\title{\Euclid preparation}
\subtitle{ LXIII. Simulations and non-linearities beyond Lambda cold dark matter. 2. Results from non-standard simulations}    
    \titlerunning{Simulations and non-linearities beyond $\Lambda$CDM. 2. Results from non-standard simulations}

\newcommand{\orcid}[1]{\orcidlink{#1}} 
\author{Euclid Collaboration: G.~R\'acz\orcid{0000-0003-3906-5699}\thanks{\email{gabor.racz@helsinki.fi}}\inst{\ref{aff1},\ref{aff2}}
\and M.-A.~Breton\inst{\ref{aff3},\ref{aff4},\ref{aff5}}
\and B.~Fiorini\orcid{0000-0002-0092-4321}\inst{\ref{aff6},\ref{aff7}}
\and A.~M.~C.~Le~Brun\orcid{0000-0002-0936-4594}\inst{\ref{aff5},\ref{aff8}}
\and H.-A.~Winther\orcid{0000-0002-6325-2710}\inst{\ref{aff9}}
\and Z.~Sakr\orcid{0000-0002-4823-3757}\inst{\ref{aff10},\ref{aff11},\ref{aff12}}
\and L.~Pizzuti\orcid{0000-0001-5654-7580}\inst{\ref{aff13}}
\and A.~Ragagnin\orcid{0000-0002-8106-2742}\inst{\ref{aff14},\ref{aff15},\ref{aff16},\ref{aff17}}
\and T.~Gayoux\orcid{0009-0008-9527-1490}\inst{\ref{aff5}}
\and E.~Altamura\orcid{0000-0001-6973-1897}\inst{\ref{aff18}}
\and E.~Carella\inst{\ref{aff19},\ref{aff20}}
\and K.~Pardede\orcid{0000-0002-7728-8220}\inst{\ref{aff21},\ref{aff22},\ref{aff23},\ref{aff15}}
\and G.~Verza\orcid{0000-0002-1886-8348}\inst{\ref{aff24},\ref{aff25}}
\and K.~Koyama\orcid{0000-0001-6727-6915}\inst{\ref{aff6}}
\and M.~Baldi\orcid{0000-0003-4145-1943}\inst{\ref{aff26},\ref{aff14},\ref{aff27}}
\and A.~Pourtsidou\orcid{0000-0001-9110-5550}\inst{\ref{aff28},\ref{aff29}}
\and F.~Vernizzi\orcid{0000-0003-3426-2802}\inst{\ref{aff30}}
\and A.~G.~Adame\orcid{0009-0005-0594-9391}\inst{\ref{aff31},\ref{aff32},\ref{aff33}}
\and J.~Adamek\orcid{0000-0002-0723-6740}\inst{\ref{aff34}}
\and S.~Avila\orcid{0000-0001-5043-3662}\inst{\ref{aff35}}
\and C.~Carbone\orcid{0000-0003-0125-3563}\inst{\ref{aff19}}
\and G.~Despali\orcid{0000-0001-6150-4112}\inst{\ref{aff16},\ref{aff14},\ref{aff27}}
\and C.~Giocoli\orcid{0000-0002-9590-7961}\inst{\ref{aff14},\ref{aff36}}
\and C.~Hern\'andez-Aguayo\orcid{0000-0001-9921-8832}\inst{\ref{aff37}}
\and F.~Hassani\orcid{0000-0003-2640-4460}\inst{\ref{aff9}}
\and M.~Kunz\orcid{0000-0002-3052-7394}\inst{\ref{aff38}}
\and B.~Li\orcid{0000-0002-1098-9188}\inst{\ref{aff39}}
\and Y.~Rasera\orcid{0000-0003-3424-6941}\inst{\ref{aff5},\ref{aff40}}
\and G.~Yepes\orcid{0000-0001-5031-7936}\inst{\ref{aff31},\ref{aff33}}
\and V.~Gonzalez-Perez\orcid{0000-0001-9938-2755}\inst{\ref{aff31}}
\and P.-S.~Corasaniti\orcid{0000-0002-6386-7846}\inst{\ref{aff5}}
\and J.~Garc\'ia-Bellido\orcid{0000-0002-9370-8360}\inst{\ref{aff32}}
\and N.~Hamaus\orcid{0000-0002-0876-2101}\inst{\ref{aff41},\ref{aff42}}
\and A.~Kiessling\orcid{0000-0002-2590-1273}\inst{\ref{aff1}}
\and M.~Marinucci\orcid{0000-0003-1159-3756}\inst{\ref{aff43},\ref{aff44}}
\and C.~Moretti\orcid{0000-0003-3314-8936}\inst{\ref{aff22},\ref{aff17},\ref{aff45},\ref{aff15},\ref{aff46}}
\and D.~F.~Mota\orcid{0000-0003-3141-142X}\inst{\ref{aff9}}
\and L.~Piga\orcid{0000-0003-2221-7406}\inst{\ref{aff47},\ref{aff21},\ref{aff19}}
\and A.~Pisani\orcid{0000-0002-6146-4437}\inst{\ref{aff48},\ref{aff49},\ref{aff25},\ref{aff50}}
\and I.~Szapudi\orcid{0000-0003-2274-0301}\inst{\ref{aff51}}
\and P.~Tallada-Cresp\'{i}\orcid{0000-0002-1336-8328}\inst{\ref{aff52},\ref{aff53}}
\and N.~Aghanim\orcid{0000-0002-6688-8992}\inst{\ref{aff54}}
\and S.~Andreon\orcid{0000-0002-2041-8784}\inst{\ref{aff55}}
\and C.~Baccigalupi\orcid{0000-0002-8211-1630}\inst{\ref{aff15},\ref{aff45},\ref{aff46},\ref{aff22}}
\and S.~Bardelli\orcid{0000-0002-8900-0298}\inst{\ref{aff14}}
\and D.~Bonino\orcid{0000-0002-3336-9977}\inst{\ref{aff56}}
\and E.~Branchini\orcid{0000-0002-0808-6908}\inst{\ref{aff57},\ref{aff58},\ref{aff55}}
\and M.~Brescia\orcid{0000-0001-9506-5680}\inst{\ref{aff59},\ref{aff60},\ref{aff61}}
\and J.~Brinchmann\orcid{0000-0003-4359-8797}\inst{\ref{aff62},\ref{aff63}}
\and S.~Camera\orcid{0000-0003-3399-3574}\inst{\ref{aff64},\ref{aff65},\ref{aff56}}
\and V.~Capobianco\orcid{0000-0002-3309-7692}\inst{\ref{aff56}}
\and V.~F.~Cardone\inst{\ref{aff66},\ref{aff67}}
\and J.~Carretero\orcid{0000-0002-3130-0204}\inst{\ref{aff52},\ref{aff53}}
\and S.~Casas\orcid{0000-0002-4751-5138}\inst{\ref{aff68}}
\and M.~Castellano\orcid{0000-0001-9875-8263}\inst{\ref{aff66}}
\and G.~Castignani\orcid{0000-0001-6831-0687}\inst{\ref{aff14}}
\and S.~Cavuoti\orcid{0000-0002-3787-4196}\inst{\ref{aff60},\ref{aff61}}
\and A.~Cimatti\inst{\ref{aff69}}
\and C.~Colodro-Conde\inst{\ref{aff70}}
\and G.~Congedo\orcid{0000-0003-2508-0046}\inst{\ref{aff28}}
\and C.~J.~Conselice\orcid{0000-0003-1949-7638}\inst{\ref{aff18}}
\and L.~Conversi\orcid{0000-0002-6710-8476}\inst{\ref{aff71},\ref{aff72}}
\and Y.~Copin\orcid{0000-0002-5317-7518}\inst{\ref{aff73}}
\and F.~Courbin\orcid{0000-0003-0758-6510}\inst{\ref{aff74}}
\and H.~M.~Courtois\orcid{0000-0003-0509-1776}\inst{\ref{aff75}}
\and A.~Da~Silva\orcid{0000-0002-6385-1609}\inst{\ref{aff76},\ref{aff77}}
\and H.~Degaudenzi\orcid{0000-0002-5887-6799}\inst{\ref{aff78}}
\and G.~De~Lucia\orcid{0000-0002-6220-9104}\inst{\ref{aff45}}
\and M.~Douspis\orcid{0000-0003-4203-3954}\inst{\ref{aff54}}
\and F.~Dubath\orcid{0000-0002-6533-2810}\inst{\ref{aff78}}
\and C.~A.~J.~Duncan\inst{\ref{aff18}}
\and X.~Dupac\inst{\ref{aff72}}
\and S.~Dusini\orcid{0000-0002-1128-0664}\inst{\ref{aff44}}
\and A.~Ealet\orcid{0000-0003-3070-014X}\inst{\ref{aff73}}
\and M.~Farina\orcid{0000-0002-3089-7846}\inst{\ref{aff79}}
\and S.~Farrens\orcid{0000-0002-9594-9387}\inst{\ref{aff80}}
\and S.~Ferriol\inst{\ref{aff73}}
\and P.~Fosalba\orcid{0000-0002-1510-5214}\inst{\ref{aff81},\ref{aff4}}
\and M.~Frailis\orcid{0000-0002-7400-2135}\inst{\ref{aff45}}
\and E.~Franceschi\orcid{0000-0002-0585-6591}\inst{\ref{aff14}}
\and M.~Fumana\orcid{0000-0001-6787-5950}\inst{\ref{aff19}}
\and S.~Galeotta\orcid{0000-0002-3748-5115}\inst{\ref{aff45}}
\and B.~Gillis\orcid{0000-0002-4478-1270}\inst{\ref{aff28}}
\and P.~G\'omez-Alvarez\orcid{0000-0002-8594-5358}\inst{\ref{aff82},\ref{aff72}}
\and A.~Grazian\orcid{0000-0002-5688-0663}\inst{\ref{aff83}}
\and F.~Grupp\inst{\ref{aff84},\ref{aff41}}
\and S.~V.~H.~Haugan\orcid{0000-0001-9648-7260}\inst{\ref{aff9}}
\and W.~Holmes\inst{\ref{aff1}}
\and F.~Hormuth\inst{\ref{aff85}}
\and A.~Hornstrup\orcid{0000-0002-3363-0936}\inst{\ref{aff86},\ref{aff87}}
\and S.~Ili\'c\orcid{0000-0003-4285-9086}\inst{\ref{aff88},\ref{aff11}}
\and K.~Jahnke\orcid{0000-0003-3804-2137}\inst{\ref{aff89}}
\and M.~Jhabvala\inst{\ref{aff90}}
\and B.~Joachimi\orcid{0000-0001-7494-1303}\inst{\ref{aff91}}
\and E.~Keih\"anen\orcid{0000-0003-1804-7715}\inst{\ref{aff92}}
\and S.~Kermiche\orcid{0000-0002-0302-5735}\inst{\ref{aff48}}
\and M.~Kilbinger\orcid{0000-0001-9513-7138}\inst{\ref{aff80}}
\and T.~Kitching\orcid{0000-0002-4061-4598}\inst{\ref{aff93}}
\and B.~Kubik\orcid{0009-0006-5823-4880}\inst{\ref{aff73}}
\and H.~Kurki-Suonio\orcid{0000-0002-4618-3063}\inst{\ref{aff2},\ref{aff94}}
\and P.~B.~Lilje\orcid{0000-0003-4324-7794}\inst{\ref{aff9}}
\and V.~Lindholm\orcid{0000-0003-2317-5471}\inst{\ref{aff2},\ref{aff94}}
\and I.~Lloro\inst{\ref{aff95}}
\and G.~Mainetti\orcid{0000-0003-2384-2377}\inst{\ref{aff96}}
\and E.~Maiorano\orcid{0000-0003-2593-4355}\inst{\ref{aff14}}
\and O.~Mansutti\orcid{0000-0001-5758-4658}\inst{\ref{aff45}}
\and O.~Marggraf\orcid{0000-0001-7242-3852}\inst{\ref{aff97}}
\and K.~Markovic\orcid{0000-0001-6764-073X}\inst{\ref{aff1}}
\and M.~Martinelli\orcid{0000-0002-6943-7732}\inst{\ref{aff66},\ref{aff67}}
\and N.~Martinet\orcid{0000-0003-2786-7790}\inst{\ref{aff98}}
\and F.~Marulli\orcid{0000-0002-8850-0303}\inst{\ref{aff16},\ref{aff14},\ref{aff27}}
\and R.~Massey\orcid{0000-0002-6085-3780}\inst{\ref{aff39}}
\and E.~Medinaceli\orcid{0000-0002-4040-7783}\inst{\ref{aff14}}
\and S.~Mei\orcid{0000-0002-2849-559X}\inst{\ref{aff99}}
\and Y.~Mellier\inst{\ref{aff100},\ref{aff8}}
\and M.~Meneghetti\orcid{0000-0003-1225-7084}\inst{\ref{aff14},\ref{aff27}}
\and G.~Meylan\inst{\ref{aff74}}
\and M.~Moresco\orcid{0000-0002-7616-7136}\inst{\ref{aff16},\ref{aff14}}
\and L.~Moscardini\orcid{0000-0002-3473-6716}\inst{\ref{aff16},\ref{aff14},\ref{aff27}}
\and S.-M.~Niemi\inst{\ref{aff101}}
\and C.~Padilla\orcid{0000-0001-7951-0166}\inst{\ref{aff35}}
\and S.~Paltani\orcid{0000-0002-8108-9179}\inst{\ref{aff78}}
\and F.~Pasian\orcid{0000-0002-4869-3227}\inst{\ref{aff45}}
\and K.~Pedersen\inst{\ref{aff102}}
\and W.~J.~Percival\orcid{0000-0002-0644-5727}\inst{\ref{aff103},\ref{aff104},\ref{aff105}}
\and V.~Pettorino\inst{\ref{aff101}}
\and S.~Pires\orcid{0000-0002-0249-2104}\inst{\ref{aff80}}
\and G.~Polenta\orcid{0000-0003-4067-9196}\inst{\ref{aff106}}
\and M.~Poncet\inst{\ref{aff107}}
\and L.~A.~Popa\inst{\ref{aff108}}
\and F.~Raison\orcid{0000-0002-7819-6918}\inst{\ref{aff84}}
\and R.~Rebolo\inst{\ref{aff70},\ref{aff109}}
\and A.~Renzi\orcid{0000-0001-9856-1970}\inst{\ref{aff43},\ref{aff44}}
\and J.~Rhodes\orcid{0000-0002-4485-8549}\inst{\ref{aff1}}
\and G.~Riccio\inst{\ref{aff60}}
\and E.~Romelli\orcid{0000-0003-3069-9222}\inst{\ref{aff45}}
\and M.~Roncarelli\orcid{0000-0001-9587-7822}\inst{\ref{aff14}}
\and R.~Saglia\orcid{0000-0003-0378-7032}\inst{\ref{aff41},\ref{aff84}}
\and J.-C.~Salvignol\inst{\ref{aff101}}
\and A.~G.~S\'anchez\orcid{0000-0003-1198-831X}\inst{\ref{aff84}}
\and D.~Sapone\orcid{0000-0001-7089-4503}\inst{\ref{aff110}}
\and B.~Sartoris\orcid{0000-0003-1337-5269}\inst{\ref{aff41},\ref{aff45}}
\and M.~Schirmer\orcid{0000-0003-2568-9994}\inst{\ref{aff89}}
\and T.~Schrabback\orcid{0000-0002-6987-7834}\inst{\ref{aff111}}
\and A.~Secroun\orcid{0000-0003-0505-3710}\inst{\ref{aff48}}
\and G.~Seidel\orcid{0000-0003-2907-353X}\inst{\ref{aff89}}
\and S.~Serrano\orcid{0000-0002-0211-2861}\inst{\ref{aff81},\ref{aff112},\ref{aff3}}
\and C.~Sirignano\orcid{0000-0002-0995-7146}\inst{\ref{aff43},\ref{aff44}}
\and G.~Sirri\orcid{0000-0003-2626-2853}\inst{\ref{aff27}}
\and L.~Stanco\orcid{0000-0002-9706-5104}\inst{\ref{aff44}}
\and J.~Steinwagner\inst{\ref{aff84}}
\and A.~N.~Taylor\inst{\ref{aff28}}
\and I.~Tereno\inst{\ref{aff76},\ref{aff113}}
\and R.~Toledo-Moreo\orcid{0000-0002-2997-4859}\inst{\ref{aff114}}
\and F.~Torradeflot\orcid{0000-0003-1160-1517}\inst{\ref{aff53},\ref{aff52}}
\and I.~Tutusaus\orcid{0000-0002-3199-0399}\inst{\ref{aff11}}
\and L.~Valenziano\orcid{0000-0002-1170-0104}\inst{\ref{aff14},\ref{aff115}}
\and T.~Vassallo\orcid{0000-0001-6512-6358}\inst{\ref{aff41},\ref{aff45}}
\and G.~Verdoes~Kleijn\orcid{0000-0001-5803-2580}\inst{\ref{aff116}}
\and Y.~Wang\orcid{0000-0002-4749-2984}\inst{\ref{aff117}}
\and J.~Weller\orcid{0000-0002-8282-2010}\inst{\ref{aff41},\ref{aff84}}
\and E.~Zucca\orcid{0000-0002-5845-8132}\inst{\ref{aff14}}
\and A.~Biviano\orcid{0000-0002-0857-0732}\inst{\ref{aff45},\ref{aff15}}
\and A.~Boucaud\orcid{0000-0001-7387-2633}\inst{\ref{aff99}}
\and E.~Bozzo\orcid{0000-0002-8201-1525}\inst{\ref{aff78}}
\and C.~Burigana\orcid{0000-0002-3005-5796}\inst{\ref{aff118},\ref{aff115}}
\and M.~Calabrese\orcid{0000-0002-2637-2422}\inst{\ref{aff119},\ref{aff19}}
\and D.~Di~Ferdinando\inst{\ref{aff27}}
\and J.~A.~Escartin~Vigo\inst{\ref{aff84}}
\and G.~Fabbian\orcid{0000-0002-3255-4695}\inst{\ref{aff120},\ref{aff121},\ref{aff25}}
\and F.~Finelli\orcid{0000-0002-6694-3269}\inst{\ref{aff14},\ref{aff115}}
\and J.~Gracia-Carpio\inst{\ref{aff84}}
\and S.~Matthew\orcid{0000-0001-8448-1697}\inst{\ref{aff28}}
\and N.~Mauri\orcid{0000-0001-8196-1548}\inst{\ref{aff69},\ref{aff27}}
\and A.~Pezzotta\orcid{0000-0003-0726-2268}\inst{\ref{aff84}}
\and M.~P\"ontinen\orcid{0000-0001-5442-2530}\inst{\ref{aff2}}
\and C.~Porciani\orcid{0000-0002-7797-2508}\inst{\ref{aff97}}
\and V.~Scottez\inst{\ref{aff100},\ref{aff122}}
\and M.~Tenti\orcid{0000-0002-4254-5901}\inst{\ref{aff27}}
\and M.~Viel\orcid{0000-0002-2642-5707}\inst{\ref{aff15},\ref{aff45},\ref{aff22},\ref{aff46},\ref{aff17}}
\and M.~Wiesmann\orcid{0009-0000-8199-5860}\inst{\ref{aff9}}
\and Y.~Akrami\orcid{0000-0002-2407-7956}\inst{\ref{aff32},\ref{aff123}}
\and V.~Allevato\orcid{0000-0001-7232-5152}\inst{\ref{aff60}}
\and S.~Anselmi\orcid{0000-0002-3579-9583}\inst{\ref{aff44},\ref{aff43},\ref{aff5}}
\and M.~Archidiacono\orcid{0000-0003-4952-9012}\inst{\ref{aff20},\ref{aff124}}
\and F.~Atrio-Barandela\orcid{0000-0002-2130-2513}\inst{\ref{aff125}}
\and A.~Balaguera-Antolinez\orcid{0000-0001-5028-3035}\inst{\ref{aff70},\ref{aff109}}
\and M.~Ballardini\orcid{0000-0003-4481-3559}\inst{\ref{aff126},\ref{aff14},\ref{aff127}}
\and D.~Bertacca\orcid{0000-0002-2490-7139}\inst{\ref{aff43},\ref{aff83},\ref{aff44}}
\and L.~Blot\orcid{0000-0002-9622-7167}\inst{\ref{aff128},\ref{aff5}}
\and S.~Borgani\orcid{0000-0001-6151-6439}\inst{\ref{aff129},\ref{aff15},\ref{aff45},\ref{aff46}}
\and S.~Bruton\orcid{0000-0002-6503-5218}\inst{\ref{aff130}}
\and R.~Cabanac\orcid{0000-0001-6679-2600}\inst{\ref{aff11}}
\and A.~Calabro\orcid{0000-0003-2536-1614}\inst{\ref{aff66}}
\and B.~Camacho~Quevedo\orcid{0000-0002-8789-4232}\inst{\ref{aff81},\ref{aff3}}
\and A.~Cappi\inst{\ref{aff14},\ref{aff131}}
\and F.~Caro\inst{\ref{aff66}}
\and C.~S.~Carvalho\inst{\ref{aff113}}
\and T.~Castro\orcid{0000-0002-6292-3228}\inst{\ref{aff45},\ref{aff46},\ref{aff15},\ref{aff17}}
\and K.~C.~Chambers\orcid{0000-0001-6965-7789}\inst{\ref{aff51}}
\and S.~Contarini\orcid{0000-0002-9843-723X}\inst{\ref{aff84}}
\and A.~R.~Cooray\orcid{0000-0002-3892-0190}\inst{\ref{aff132}}
\and B.~De~Caro\inst{\ref{aff19}}
\and S.~de~la~Torre\inst{\ref{aff98}}
\and G.~Desprez\inst{\ref{aff133}}
\and A.~D\'iaz-S\'anchez\orcid{0000-0003-0748-4768}\inst{\ref{aff134}}
\and J.~J.~Diaz\inst{\ref{aff135}}
\and S.~Di~Domizio\orcid{0000-0003-2863-5895}\inst{\ref{aff57},\ref{aff58}}
\and H.~Dole\orcid{0000-0002-9767-3839}\inst{\ref{aff54}}
\and S.~Escoffier\orcid{0000-0002-2847-7498}\inst{\ref{aff48}}
\and A.~G.~Ferrari\orcid{0009-0005-5266-4110}\inst{\ref{aff69},\ref{aff27}}
\and P.~G.~Ferreira\orcid{0000-0002-3021-2851}\inst{\ref{aff136}}
\and I.~Ferrero\orcid{0000-0002-1295-1132}\inst{\ref{aff9}}
\and A.~Fontana\orcid{0000-0003-3820-2823}\inst{\ref{aff66}}
\and F.~Fornari\orcid{0000-0003-2979-6738}\inst{\ref{aff115}}
\and L.~Gabarra\orcid{0000-0002-8486-8856}\inst{\ref{aff136}}
\and K.~Ganga\orcid{0000-0001-8159-8208}\inst{\ref{aff99}}
\and T.~Gasparetto\orcid{0000-0002-7913-4866}\inst{\ref{aff45}}
\and E.~Gaztanaga\orcid{0000-0001-9632-0815}\inst{\ref{aff3},\ref{aff81},\ref{aff6}}
\and F.~Giacomini\orcid{0000-0002-3129-2814}\inst{\ref{aff27}}
\and F.~Gianotti\orcid{0000-0003-4666-119X}\inst{\ref{aff14}}
\and G.~Gozaliasl\orcid{0000-0002-0236-919X}\inst{\ref{aff137},\ref{aff2}}
\and C.~M.~Gutierrez\orcid{0000-0001-7854-783X}\inst{\ref{aff138}}
\and A.~Hall\orcid{0000-0002-3139-8651}\inst{\ref{aff28}}
\and H.~Hildebrandt\orcid{0000-0002-9814-3338}\inst{\ref{aff139}}
\and J.~Hjorth\orcid{0000-0002-4571-2306}\inst{\ref{aff140}}
\and A.~Jimenez~Mu\~noz\orcid{0009-0004-5252-185X}\inst{\ref{aff141}}
\and J.~J.~E.~Kajava\orcid{0000-0002-3010-8333}\inst{\ref{aff142},\ref{aff143}}
\and V.~Kansal\orcid{0000-0002-4008-6078}\inst{\ref{aff144},\ref{aff145}}
\and D.~Karagiannis\orcid{0000-0002-4927-0816}\inst{\ref{aff7},\ref{aff146}}
\and C.~C.~Kirkpatrick\inst{\ref{aff92}}
\and F.~Lacasa\orcid{0000-0002-7268-3440}\inst{\ref{aff147},\ref{aff54}}
\and J.~Le~Graet\orcid{0000-0001-6523-7971}\inst{\ref{aff48}}
\and L.~Legrand\orcid{0000-0003-0610-5252}\inst{\ref{aff148}}
\and J.~Lesgourgues\orcid{0000-0001-7627-353X}\inst{\ref{aff68}}
\and T.~I.~Liaudat\orcid{0000-0002-9104-314X}\inst{\ref{aff149}}
\and A.~Loureiro\orcid{0000-0002-4371-0876}\inst{\ref{aff150},\ref{aff151}}
\and J.~Macias-Perez\orcid{0000-0002-5385-2763}\inst{\ref{aff141}}
\and G.~Maggio\orcid{0000-0003-4020-4836}\inst{\ref{aff45}}
\and M.~Magliocchetti\orcid{0000-0001-9158-4838}\inst{\ref{aff79}}
\and F.~Mannucci\orcid{0000-0002-4803-2381}\inst{\ref{aff152}}
\and R.~Maoli\orcid{0000-0002-6065-3025}\inst{\ref{aff153},\ref{aff66}}
\and C.~J.~A.~P.~Martins\orcid{0000-0002-4886-9261}\inst{\ref{aff154},\ref{aff62}}
\and L.~Maurin\orcid{0000-0002-8406-0857}\inst{\ref{aff54}}
\and R.~B.~Metcalf\orcid{0000-0003-3167-2574}\inst{\ref{aff16},\ref{aff14}}
\and M.~Miluzio\inst{\ref{aff72},\ref{aff155}}
\and P.~Monaco\orcid{0000-0003-2083-7564}\inst{\ref{aff129},\ref{aff45},\ref{aff46},\ref{aff15}}
\and A.~Montoro\orcid{0000-0003-4730-8590}\inst{\ref{aff3},\ref{aff81}}
\and A.~Mora\orcid{0000-0002-1922-8529}\inst{\ref{aff156}}
\and G.~Morgante\inst{\ref{aff14}}
\and S.~Nadathur\orcid{0000-0001-9070-3102}\inst{\ref{aff6}}
\and L.~Patrizii\inst{\ref{aff27}}
\and V.~Popa\orcid{0000-0002-9118-8330}\inst{\ref{aff108}}
\and D.~Potter\orcid{0000-0002-0757-5195}\inst{\ref{aff34}}
\and P.~Reimberg\orcid{0000-0003-3410-0280}\inst{\ref{aff100}}
\and I.~Risso\orcid{0000-0003-2525-7761}\inst{\ref{aff157}}
\and P.-F.~Rocci\inst{\ref{aff54}}
\and M.~Sahl\'en\orcid{0000-0003-0973-4804}\inst{\ref{aff158}}
\and A.~Schneider\orcid{0000-0001-7055-8104}\inst{\ref{aff34}}
\and M.~Sereno\orcid{0000-0003-0302-0325}\inst{\ref{aff14},\ref{aff27}}
\and A.~Silvestri\orcid{0000-0001-6904-5061}\inst{\ref{aff159}}
\and A.~Spurio~Mancini\orcid{0000-0001-5698-0990}\inst{\ref{aff160},\ref{aff93}}
\and J.~Stadel\orcid{0000-0001-7565-8622}\inst{\ref{aff34}}
\and K.~Tanidis\inst{\ref{aff136}}
\and C.~Tao\orcid{0000-0001-7961-8177}\inst{\ref{aff48}}
\and N.~Tessore\orcid{0000-0002-9696-7931}\inst{\ref{aff91}}
\and G.~Testera\inst{\ref{aff58}}
\and R.~Teyssier\orcid{0000-0001-7689-0933}\inst{\ref{aff49}}
\and S.~Toft\orcid{0000-0003-3631-7176}\inst{\ref{aff87},\ref{aff161},\ref{aff162}}
\and S.~Tosi\orcid{0000-0002-7275-9193}\inst{\ref{aff57},\ref{aff58}}
\and A.~Troja\orcid{0000-0003-0239-4595}\inst{\ref{aff43},\ref{aff44}}
\and M.~Tucci\inst{\ref{aff78}}
\and C.~Valieri\inst{\ref{aff27}}
\and J.~Valiviita\orcid{0000-0001-6225-3693}\inst{\ref{aff2},\ref{aff94}}
\and D.~Vergani\orcid{0000-0003-0898-2216}\inst{\ref{aff14}}
\and P.~Vielzeuf\orcid{0000-0003-2035-9339}\inst{\ref{aff48}}
\and N.~A.~Walton\orcid{0000-0003-3983-8778}\inst{\ref{aff120}}}
										   
\institute{Jet Propulsion Laboratory, California Institute of Technology, 4800 Oak Grove Drive, Pasadena, CA, 91109, USA\label{aff1}
\and
Department of Physics, P.O. Box 64, 00014 University of Helsinki, Finland\label{aff2}
\and
Institute of Space Sciences (ICE, CSIC), Campus UAB, Carrer de Can Magrans, s/n, 08193 Barcelona, Spain\label{aff3}
\and
Institut de Ciencies de l'Espai (IEEC-CSIC), Campus UAB, Carrer de Can Magrans, s/n Cerdanyola del Vall\'es, 08193 Barcelona, Spain\label{aff4}
\and
Laboratoire Univers et Th\'eorie, Observatoire de Paris, Universit\'e PSL, Universit\'e Paris Cit\'e, CNRS, 92190 Meudon, France\label{aff5}
\and
Institute of Cosmology and Gravitation, University of Portsmouth, Portsmouth PO1 3FX, UK\label{aff6}
\and
School of Physics and Astronomy, Queen Mary University of London, Mile End Road, London E1 4NS, UK\label{aff7}
\and
Institut d'Astrophysique de Paris, UMR 7095, CNRS, and Sorbonne Universit\'e, 98 bis boulevard Arago, 75014 Paris, France\label{aff8}
\and
Institute of Theoretical Astrophysics, University of Oslo, P.O. Box 1029 Blindern, 0315 Oslo, Norway\label{aff9}
\and
Institut f\"ur Theoretische Physik, University of Heidelberg, Philosophenweg 16, 69120 Heidelberg, Germany\label{aff10}
\and
Institut de Recherche en Astrophysique et Plan\'etologie (IRAP), Universit\'e de Toulouse, CNRS, UPS, CNES, 14 Av. Edouard Belin, 31400 Toulouse, France\label{aff11}
\and
Universit\'e St Joseph; Faculty of Sciences, Beirut, Lebanon\label{aff12}
\and
Dipartimento di Fisica ``G. Occhialini", Universit\`a degli Studi di Milano Bicocca, Piazza della Scienza 3, 20126 Milano, Italy\label{aff13}
\and
INAF-Osservatorio di Astrofisica e Scienza dello Spazio di Bologna, Via Piero Gobetti 93/3, 40129 Bologna, Italy\label{aff14}
\and
IFPU, Institute for Fundamental Physics of the Universe, via Beirut 2, 34151 Trieste, Italy\label{aff15}
\and
Dipartimento di Fisica e Astronomia "Augusto Righi" - Alma Mater Studiorum Universit\`a di Bologna, via Piero Gobetti 93/2, 40129 Bologna, Italy\label{aff16}
\and
ICSC - Centro Nazionale di Ricerca in High Performance Computing, Big Data e Quantum Computing, Via Magnanelli 2, Bologna, Italy\label{aff17}
\and
Jodrell Bank Centre for Astrophysics, Department of Physics and Astronomy, University of Manchester, Oxford Road, Manchester M13 9PL, UK\label{aff18}
\and
INAF-IASF Milano, Via Alfonso Corti 12, 20133 Milano, Italy\label{aff19}
\and
Dipartimento di Fisica "Aldo Pontremoli", Universit\`a degli Studi di Milano, Via Celoria 16, 20133 Milano, Italy\label{aff20}
\and
INFN Gruppo Collegato di Parma, Viale delle Scienze 7/A 43124 Parma, Italy\label{aff21}
\and
SISSA, International School for Advanced Studies, Via Bonomea 265, 34136 Trieste TS, Italy\label{aff22}
\and
International Centre for Theoretical Physics (ICTP), Strada Costiera 11, 34151 Trieste, Italy\label{aff23}
\and
Center for Cosmology and Particle Physics, Department of Physics, New York University, New York, NY 10003, USA\label{aff24}
\and
Center for Computational Astrophysics, Flatiron Institute, 162 5th Avenue, 10010, New York, NY, USA\label{aff25}
\and
Dipartimento di Fisica e Astronomia, Universit\`a di Bologna, Via Gobetti 93/2, 40129 Bologna, Italy\label{aff26}
\and
INFN-Sezione di Bologna, Viale Berti Pichat 6/2, 40127 Bologna, Italy\label{aff27}
\and
Institute for Astronomy, University of Edinburgh, Royal Observatory, Blackford Hill, Edinburgh EH9 3HJ, UK\label{aff28}
\and
Higgs Centre for Theoretical Physics, School of Physics and Astronomy, The University of Edinburgh, Edinburgh EH9 3FD, UK\label{aff29}
\and
Institut de Physique Th\'eorique, CEA, CNRS, Universit\'e Paris-Saclay 91191 Gif-sur-Yvette Cedex, France\label{aff30}
\and
Departamento de F\'isica Te\'orica, Facultad de Ciencias, Universidad Aut\'onoma de Madrid, 28049 Cantoblanco, Madrid, Spain\label{aff31}
\and
Instituto de F\'isica Te\'orica UAM-CSIC, Campus de Cantoblanco, 28049 Madrid, Spain\label{aff32}
\and
Centro de Investigaci\'{o}n Avanzada en F\'isica Fundamental (CIAFF), Facultad de Ciencias, Universidad Aut\'{o}noma de Madrid, 28049 Madrid, Spain\label{aff33}
\and
Department of Astrophysics, University of Zurich, Winterthurerstrasse 190, 8057 Zurich, Switzerland\label{aff34}
\and
Institut de F\'{i}sica d'Altes Energies (IFAE), The Barcelona Institute of Science and Technology, Campus UAB, 08193 Bellaterra (Barcelona), Spain\label{aff35}
\and
Istituto Nazionale di Fisica Nucleare, Sezione di Bologna, Via Irnerio 46, 40126 Bologna, Italy\label{aff36}
\and
Max-Planck-Institut f\"ur Astrophysik, Karl-Schwarzschild-Str.~1, 85748 Garching, Germany\label{aff37}
\and
Universit\'e de Gen\`eve, D\'epartement de Physique Th\'eorique and Centre for Astroparticle Physics, 24 quai Ernest-Ansermet, CH-1211 Gen\`eve 4, Switzerland\label{aff38}
\and
Department of Physics, Institute for Computational Cosmology, Durham University, South Road, DH1 3LE, UK\label{aff39}
\and
Institut universitaire de France (IUF), 1 rue Descartes, 75231 PARIS CEDEX 05, France\label{aff40}
\and
Universit\"ats-Sternwarte M\"unchen, Fakult\"at f\"ur Physik, Ludwig-Maximilians-Universit\"at M\"unchen, Scheinerstrasse 1, 81679 M\"unchen, Germany\label{aff41}
\and
Excellence Cluster ORIGINS, Boltzmannstrasse 2, 85748 Garching, Germany\label{aff42}
\and
Dipartimento di Fisica e Astronomia "G. Galilei", Universit\`a di Padova, Via Marzolo 8, 35131 Padova, Italy\label{aff43}
\and
INFN-Padova, Via Marzolo 8, 35131 Padova, Italy\label{aff44}
\and
INAF-Osservatorio Astronomico di Trieste, Via G. B. Tiepolo 11, 34143 Trieste, Italy\label{aff45}
\and
INFN, Sezione di Trieste, Via Valerio 2, 34127 Trieste TS, Italy\label{aff46}
\and
Dipartimento di Scienze Matematiche, Fisiche e Informatiche, Universit\`a di Parma, Viale delle Scienze 7/A 43124 Parma, Italy\label{aff47}
\and
Aix-Marseille Universit\'e, CNRS/IN2P3, CPPM, Marseille, France\label{aff48}
\and
Department of Astrophysical Sciences, Peyton Hall, Princeton University, Princeton, NJ 08544, USA\label{aff49}
\and
The Cooper Union for the Advancement of Science and Art, 41 Cooper Square, New York, NY 10003, USA\label{aff50}
\and
Institute for Astronomy, University of Hawaii, 2680 Woodlawn Drive, Honolulu, HI 96822, USA\label{aff51}
\and
Centro de Investigaciones Energ\'eticas, Medioambientales y Tecnol\'ogicas (CIEMAT), Avenida Complutense 40, 28040 Madrid, Spain\label{aff52}
\and
Port d'Informaci\'{o} Cient\'{i}fica, Campus UAB, C. Albareda s/n, 08193 Bellaterra (Barcelona), Spain\label{aff53}
\and
Universit\'e Paris-Saclay, CNRS, Institut d'astrophysique spatiale, 91405, Orsay, France\label{aff54}
\and
INAF-Osservatorio Astronomico di Brera, Via Brera 28, 20122 Milano, Italy\label{aff55}
\and
INAF-Osservatorio Astrofisico di Torino, Via Osservatorio 20, 10025 Pino Torinese (TO), Italy\label{aff56}
\and
Dipartimento di Fisica, Universit\`a di Genova, Via Dodecaneso 33, 16146, Genova, Italy\label{aff57}
\and
INFN-Sezione di Genova, Via Dodecaneso 33, 16146, Genova, Italy\label{aff58}
\and
Department of Physics "E. Pancini", University Federico II, Via Cinthia 6, 80126, Napoli, Italy\label{aff59}
\and
INAF-Osservatorio Astronomico di Capodimonte, Via Moiariello 16, 80131 Napoli, Italy\label{aff60}
\and
INFN section of Naples, Via Cinthia 6, 80126, Napoli, Italy\label{aff61}
\and
Instituto de Astrof\'isica e Ci\^encias do Espa\c{c}o, Universidade do Porto, CAUP, Rua das Estrelas, PT4150-762 Porto, Portugal\label{aff62}
\and
Faculdade de Ci\^encias da Universidade do Porto, Rua do Campo de Alegre, 4150-007 Porto, Portugal\label{aff63}
\and
Dipartimento di Fisica, Universit\`a degli Studi di Torino, Via P. Giuria 1, 10125 Torino, Italy\label{aff64}
\and
INFN-Sezione di Torino, Via P. Giuria 1, 10125 Torino, Italy\label{aff65}
\and
INAF-Osservatorio Astronomico di Roma, Via Frascati 33, 00078 Monteporzio Catone, Italy\label{aff66}
\and
INFN-Sezione di Roma, Piazzale Aldo Moro, 2 - c/o Dipartimento di Fisica, Edificio G. Marconi, 00185 Roma, Italy\label{aff67}
\and
Institute for Theoretical Particle Physics and Cosmology (TTK), RWTH Aachen University, 52056 Aachen, Germany\label{aff68}
\and
Dipartimento di Fisica e Astronomia "Augusto Righi" - Alma Mater Studiorum Universit\`a di Bologna, Viale Berti Pichat 6/2, 40127 Bologna, Italy\label{aff69}
\and
Instituto de Astrof\'isica de Canarias, Calle V\'ia L\'actea s/n, 38204, San Crist\'obal de La Laguna, Tenerife, Spain\label{aff70}
\and
European Space Agency/ESRIN, Largo Galileo Galilei 1, 00044 Frascati, Roma, Italy\label{aff71}
\and
ESAC/ESA, Camino Bajo del Castillo, s/n., Urb. Villafranca del Castillo, 28692 Villanueva de la Ca\~nada, Madrid, Spain\label{aff72}
\and
Universit\'e Claude Bernard Lyon 1, CNRS/IN2P3, IP2I Lyon, UMR 5822, Villeurbanne, F-69100, France\label{aff73}
\and
Institute of Physics, Laboratory of Astrophysics, Ecole Polytechnique F\'ed\'erale de Lausanne (EPFL), Observatoire de Sauverny, 1290 Versoix, Switzerland\label{aff74}
\and
UCB Lyon 1, CNRS/IN2P3, IUF, IP2I Lyon, 4 rue Enrico Fermi, 69622 Villeurbanne, France\label{aff75}
\and
Departamento de F\'isica, Faculdade de Ci\^encias, Universidade de Lisboa, Edif\'icio C8, Campo Grande, PT1749-016 Lisboa, Portugal\label{aff76}
\and
Instituto de Astrof\'isica e Ci\^encias do Espa\c{c}o, Faculdade de Ci\^encias, Universidade de Lisboa, Campo Grande, 1749-016 Lisboa, Portugal\label{aff77}
\and
Department of Astronomy, University of Geneva, ch. d'Ecogia 16, 1290 Versoix, Switzerland\label{aff78}
\and
INAF-Istituto di Astrofisica e Planetologia Spaziali, via del Fosso del Cavaliere, 100, 00100 Roma, Italy\label{aff79}
\and
Universit\'e Paris-Saclay, Universit\'e Paris Cit\'e, CEA, CNRS, AIM, 91191, Gif-sur-Yvette, France\label{aff80}
\and
Institut d'Estudis Espacials de Catalunya (IEEC),  Edifici RDIT, Campus UPC, 08860 Castelldefels, Barcelona, Spain\label{aff81}
\and
FRACTAL S.L.N.E., calle Tulip\'an 2, Portal 13 1A, 28231, Las Rozas de Madrid, Spain\label{aff82}
\and
INAF-Osservatorio Astronomico di Padova, Via dell'Osservatorio 5, 35122 Padova, Italy\label{aff83}
\and
Max Planck Institute for Extraterrestrial Physics, Giessenbachstr. 1, 85748 Garching, Germany\label{aff84}
\and
Felix Hormuth Engineering, Goethestr. 17, 69181 Leimen, Germany\label{aff85}
\and
Technical University of Denmark, Elektrovej 327, 2800 Kgs. Lyngby, Denmark\label{aff86}
\and
Cosmic Dawn Center (DAWN), Denmark\label{aff87}
\and
Universit\'e Paris-Saclay, CNRS/IN2P3, IJCLab, 91405 Orsay, France\label{aff88}
\and
Max-Planck-Institut f\"ur Astronomie, K\"onigstuhl 17, 69117 Heidelberg, Germany\label{aff89}
\and
NASA Goddard Space Flight Center, Greenbelt, MD 20771, USA\label{aff90}
\and
Department of Physics and Astronomy, University College London, Gower Street, London WC1E 6BT, UK\label{aff91}
\and
Department of Physics and Helsinki Institute of Physics, Gustaf H\"allstr\"omin katu 2, 00014 University of Helsinki, Finland\label{aff92}
\and
Mullard Space Science Laboratory, University College London, Holmbury St Mary, Dorking, Surrey RH5 6NT, UK\label{aff93}
\and
Helsinki Institute of Physics, Gustaf H{\"a}llstr{\"o}min katu 2, University of Helsinki, Helsinki, Finland\label{aff94}
\and
NOVA optical infrared instrumentation group at ASTRON, Oude Hoogeveensedijk 4, 7991PD, Dwingeloo, The Netherlands\label{aff95}
\and
Centre de Calcul de l'IN2P3/CNRS, 21 avenue Pierre de Coubertin 69627 Villeurbanne Cedex, France\label{aff96}
\and
Universit\"at Bonn, Argelander-Institut f\"ur Astronomie, Auf dem H\"ugel 71, 53121 Bonn, Germany\label{aff97}
\and
Aix-Marseille Universit\'e, CNRS, CNES, LAM, Marseille, France\label{aff98}
\and
Universit\'e Paris Cit\'e, CNRS, Astroparticule et Cosmologie, 75013 Paris, France\label{aff99}
\and
Institut d'Astrophysique de Paris, 98bis Boulevard Arago, 75014, Paris, France\label{aff100}
\and
European Space Agency/ESTEC, Keplerlaan 1, 2201 AZ Noordwijk, The Netherlands\label{aff101}
\and
Department of Physics and Astronomy, University of Aarhus, Ny Munkegade 120, DK-8000 Aarhus C, Denmark\label{aff102}
\and
Waterloo Centre for Astrophysics, University of Waterloo, Waterloo, Ontario N2L 3G1, Canada\label{aff103}
\and
Department of Physics and Astronomy, University of Waterloo, Waterloo, Ontario N2L 3G1, Canada\label{aff104}
\and
Perimeter Institute for Theoretical Physics, Waterloo, Ontario N2L 2Y5, Canada\label{aff105}
\and
Space Science Data Center, Italian Space Agency, via del Politecnico snc, 00133 Roma, Italy\label{aff106}
\and
Centre National d'Etudes Spatiales -- Centre spatial de Toulouse, 18 avenue Edouard Belin, 31401 Toulouse Cedex 9, France\label{aff107}
\and
Institute of Space Science, Str. Atomistilor, nr. 409 M\u{a}gurele, Ilfov, 077125, Romania\label{aff108}
\and
Departamento de Astrof\'isica, Universidad de La Laguna, 38206, La Laguna, Tenerife, Spain\label{aff109}
\and
Departamento de F\'isica, FCFM, Universidad de Chile, Blanco Encalada 2008, Santiago, Chile\label{aff110}
\and
Universit\"at Innsbruck, Institut f\"ur Astro- und Teilchenphysik, Technikerstr. 25/8, 6020 Innsbruck, Austria\label{aff111}
\and
Satlantis, University Science Park, Sede Bld 48940, Leioa-Bilbao, Spain\label{aff112}
\and
Instituto de Astrof\'isica e Ci\^encias do Espa\c{c}o, Faculdade de Ci\^encias, Universidade de Lisboa, Tapada da Ajuda, 1349-018 Lisboa, Portugal\label{aff113}
\and
Universidad Polit\'ecnica de Cartagena, Departamento de Electr\'onica y Tecnolog\'ia de Computadoras,  Plaza del Hospital 1, 30202 Cartagena, Spain\label{aff114}
\and
INFN-Bologna, Via Irnerio 46, 40126 Bologna, Italy\label{aff115}
\and
Kapteyn Astronomical Institute, University of Groningen, PO Box 800, 9700 AV Groningen, The Netherlands\label{aff116}
\and
Infrared Processing and Analysis Center, California Institute of Technology, Pasadena, CA 91125, USA\label{aff117}
\and
INAF, Istituto di Radioastronomia, Via Piero Gobetti 101, 40129 Bologna, Italy\label{aff118}
\and
Astronomical Observatory of the Autonomous Region of the Aosta Valley (OAVdA), Loc. Lignan 39, I-11020, Nus (Aosta Valley), Italy\label{aff119}
\and
Institute of Astronomy, University of Cambridge, Madingley Road, Cambridge CB3 0HA, UK\label{aff120}
\and
School of Physics and Astronomy, Cardiff University, The Parade, Cardiff, CF24 3AA, UK\label{aff121}
\and
Junia, EPA department, 41 Bd Vauban, 59800 Lille, France\label{aff122}
\and
CERCA/ISO, Department of Physics, Case Western Reserve University, 10900 Euclid Avenue, Cleveland, OH 44106, USA\label{aff123}
\and
INFN-Sezione di Milano, Via Celoria 16, 20133 Milano, Italy\label{aff124}
\and
Departamento de F{\'\i}sica Fundamental. Universidad de Salamanca. Plaza de la Merced s/n. 37008 Salamanca, Spain\label{aff125}
\and
Dipartimento di Fisica e Scienze della Terra, Universit\`a degli Studi di Ferrara, Via Giuseppe Saragat 1, 44122 Ferrara, Italy\label{aff126}
\and
Istituto Nazionale di Fisica Nucleare, Sezione di Ferrara, Via Giuseppe Saragat 1, 44122 Ferrara, Italy\label{aff127}
\and
Kavli Institute for the Physics and Mathematics of the Universe (WPI), University of Tokyo, Kashiwa, Chiba 277-8583, Japan\label{aff128}
\and
Dipartimento di Fisica - Sezione di Astronomia, Universit\`a di Trieste, Via Tiepolo 11, 34131 Trieste, Italy\label{aff129}
\and
Minnesota Institute for Astrophysics, University of Minnesota, 116 Church St SE, Minneapolis, MN 55455, USA\label{aff130}
\and
Universit\'e C\^{o}te d'Azur, Observatoire de la C\^{o}te d'Azur, CNRS, Laboratoire Lagrange, Bd de l'Observatoire, CS 34229, 06304 Nice cedex 4, France\label{aff131}
\and
Department of Physics \& Astronomy, University of California Irvine, Irvine CA 92697, USA\label{aff132}
\and
Department of Astronomy \& Physics and Institute for Computational Astrophysics, Saint Mary's University, 923 Robie Street, Halifax, Nova Scotia, B3H 3C3, Canada\label{aff133}
\and
Departamento F\'isica Aplicada, Universidad Polit\'ecnica de Cartagena, Campus Muralla del Mar, 30202 Cartagena, Murcia, Spain\label{aff134}
\and
Instituto de Astrof\'isica de Canarias (IAC); Departamento de Astrof\'isica, Universidad de La Laguna (ULL), 38200, La Laguna, Tenerife, Spain\label{aff135}
\and
Department of Physics, Oxford University, Keble Road, Oxford OX1 3RH, UK\label{aff136}
\and
Department of Computer Science, Aalto University, PO Box 15400, Espoo, FI-00 076, Finland\label{aff137}
\and
Instituto de Astrof\'\i sica de Canarias, c/ Via Lactea s/n, La Laguna E-38200, Spain. Departamento de Astrof\'\i sica de la Universidad de La Laguna, Avda. Francisco Sanchez, La Laguna, E-38200, Spain\label{aff138}
\and
Ruhr University Bochum, Faculty of Physics and Astronomy, Astronomical Institute (AIRUB), German Centre for Cosmological Lensing (GCCL), 44780 Bochum, Germany\label{aff139}
\and
DARK, Niels Bohr Institute, University of Copenhagen, Jagtvej 155, 2200 Copenhagen, Denmark\label{aff140}
\and
Univ. Grenoble Alpes, CNRS, Grenoble INP, LPSC-IN2P3, 53, Avenue des Martyrs, 38000, Grenoble, France\label{aff141}
\and
Department of Physics and Astronomy, Vesilinnantie 5, 20014 University of Turku, Finland\label{aff142}
\and
Serco for European Space Agency (ESA), Camino bajo del Castillo, s/n, Urbanizacion Villafranca del Castillo, Villanueva de la Ca\~nada, 28692 Madrid, Spain\label{aff143}
\and
ARC Centre of Excellence for Dark Matter Particle Physics, Melbourne, Australia\label{aff144}
\and
Centre for Astrophysics \& Supercomputing, Swinburne University of Technology,  Hawthorn, Victoria 3122, Australia\label{aff145}
\and
Department of Physics and Astronomy, University of the Western Cape, Bellville, Cape Town, 7535, South Africa\label{aff146}
\and
Universit\'e Libre de Bruxelles (ULB), Service de Physique Th\'eorique CP225, Boulevard du Triophe, 1050 Bruxelles, Belgium\label{aff147}
\and
ICTP South American Institute for Fundamental Research, Instituto de F\'{\i}sica Te\'orica, Universidade Estadual Paulista, S\~ao Paulo, Brazil\label{aff148}
\and
IRFU, CEA, Universit\'e Paris-Saclay 91191 Gif-sur-Yvette Cedex, France\label{aff149}
\and
Oskar Klein Centre for Cosmoparticle Physics, Department of Physics, Stockholm University, Stockholm, SE-106 91, Sweden\label{aff150}
\and
Astrophysics Group, Blackett Laboratory, Imperial College London, London SW7 2AZ, UK\label{aff151}
\and
INAF-Osservatorio Astrofisico di Arcetri, Largo E. Fermi 5, 50125, Firenze, Italy\label{aff152}
\and
Dipartimento di Fisica, Sapienza Universit\`a di Roma, Piazzale Aldo Moro 2, 00185 Roma, Italy\label{aff153}
\and
Centro de Astrof\'{\i}sica da Universidade do Porto, Rua das Estrelas, 4150-762 Porto, Portugal\label{aff154}
\and
HE Space for European Space Agency (ESA), Camino bajo del Castillo, s/n, Urbanizacion Villafranca del Castillo, Villanueva de la Ca\~nada, 28692 Madrid, Spain\label{aff155}
\and
Aurora Technology for European Space Agency (ESA), Camino bajo del Castillo, s/n, Urbanizacion Villafranca del Castillo, Villanueva de la Ca\~nada, 28692 Madrid, Spain\label{aff156}
\and
Dipartimento di Fisica, Universit\`a degli studi di Genova, and INFN-Sezione di Genova, via Dodecaneso 33, 16146, Genova, Italy\label{aff157}
\and
Theoretical astrophysics, Department of Physics and Astronomy, Uppsala University, Box 515, 751 20 Uppsala, Sweden\label{aff158}
\and
Institute Lorentz, Leiden University, Niels Bohrweg 2, 2333 CA Leiden, The Netherlands\label{aff159}
\and
Department of Physics, Royal Holloway, University of London, TW20 0EX, UK\label{aff160}
\and
Cosmic Dawn Center (DAWN)\label{aff161}
\and
Niels Bohr Institute, University of Copenhagen, Jagtvej 128, 2200 Copenhagen, Denmark\label{aff162}}    

   \authorrunning{Euclid Collaboration: Racz et al.}
   \date{\today}

 
  \abstract{The \euclid mission will measure cosmological parameters with unprecedented precision. To distinguish between cosmological models, it is essential to generate realistic mock observables from cosmological simulations that were run in both the standard $\Lambda$-cold-dark-matter (\lcdm) paradigm and in many non-standard models beyond \lcdm. 
  We present the scientific results from a suite of cosmological {\it N}-body simulations using non-standard models including dynamical dark energy, \textit{k}-essence, interacting dark energy, modified gravity, massive neutrinos, and primordial non-Gaussianities. We investigate how these models affect the large-scale-structure formation and evolution in addition to providing synthetic observables that can be used to test and constrain these models with \euclid data.
  We developed a custom pipeline based on the \rockstar halo finder and the \nbodykit large-scale structure toolkit to analyse the particle output of non-standard simulations and generate mock observables such as halo and void catalogues, mass density fields, and power spectra in a consistent way. We compare these observables with those from the standard \lcdm model and quantify the deviations.
  We find that non-standard cosmological models can leave large imprints on the synthetic observables that we have generated. Our results demonstrate that non-standard cosmological {\it N}-body simulations provide valuable insights into the physics of dark energy and dark matter, which is essential to maximising the scientific return of \euclid.}
  

   \maketitle
%

\section{Introduction}
\blfootnote{\copyright 2024. All rights reserved.}
The concordance $\Lambda$-cold-dark-matter (\lcdm) model is the simplest cosmological scenario that accounts for the cosmological observations thus far available. It is based on the assumption that in addition to baryonic matter and radiation, the Universe is filled with two invisible components: an exotic form of matter, dubbed dark energy and described by a cosmological constant ($\Lambda$) in Einstein's equations of general relativity, and a cold-dark-matter (CDM) component that is non-relativistic and only interacts through gravity. In this scenario, dark matter is primarily responsible for fostering the formation of the visible structures we observe today, while dark energy drives the accelerated expansion of the Universe at late times. This model has been remarkably successful in explaining a variety of cosmological observations, such as the Hubble diagram from luminosity distance measurements of type Ia supernovae \citep{1998AJ....116.1009R, 1999ApJ...517..565P}, temperature and polarisation anisotropy angular power spectra of the cosmic microwave background \citep[CMB, ][]{2000Natur.404..955D,2003ApJS..148..175S, 2002Natur.420..772K,2020AA...641A...6P}, the galaxy power spectrum of the large-scale structure \citep[LSS, ][]{2002MNRAS.330L..29E,2003astro.ph..6581C, 2004ApJ...606..702T,2006PhRvD..74l3507T}, and the presence of baryonic acoustic oscillations (BAO) in the LSS \citep{2005ApJ...633..560E, 2005MNRAS.362..505C}. Despite the great success of the \lcdm model, the physical origin of dark energy and dark matter remains unknown. Unveiling the nature of these dark components is the primary motivation for many investigations in modern cosmology. 

In the last decade, multiple tensions among different types of cosmological observations have emerged. As an example, while CMB measurements indicate a value of the Hubble constant of $H_0=67.7\,\pm\,0.4\,\kmsmpc$ \citep{2020AA...641A...6P}, local measurements, often based on the observations of supernovae in nearby galaxies, suggest a higher value of $H_0=73.0\,\pm\,1.0\,\kmsmpc$ \citep{2022ApJ...934L...7R}. This $5\sigma$ discrepancy is called the Hubble tension. A similar tension has been identified in the $S_8=\sigma_8\,\sqrt{\om/0.3}$ parameter, which combines the amplitude of linear-matter density fluctuations on the $8\mpcoh$ scale, $\sigma_8$, and the cosmic matter density, $\om$. Measurements derived from the CMB \citep{2020AA...641A...6P} appear to yield a value of $S_8$ $2.9\sigma$ higher than that obtained from observations of the LSS \citep{2023PhRvD.108b3520J}, such as measurements of the clustering of galaxies and weak gravitational lensing \citep{2023arXiv230611124L,2022PhRvD.105b3520A}. Such tensions may result from systematic errors yet to be identified in the data. Alternatively, they may be a manifestation of the limits of the \lcdm model, since modifications to the standard cosmological model can provide a solution to these tensions \citep[see e.g.][]{2019Symm...11..986M, 2021CQGra..38o3001D}. 

Ongoing and upcoming Stage-IV surveys, such as \euclid \citep{EuclidSkyOverview,2011arXiv1110.3193L, 2022AA...662A.112E}, Dark Energy Spectroscopic Instrument \citep[hereafter DESI, ][]{2016arXiv161100036D}, Vera C. Rubin Observatory Legacy Survey of Space and Time \citep[LSST, ][]{2019ApJ...873..111I}, Spectro-Photometer for the History of the Universe, Epoch of Reionization, and the Ices Explorer \citep[SPHEREx, ][]{2014arXiv1412.4872D}, and the \textit{Nancy Grace Roman} Space Telescope \citep{2015arXiv150303757S}, will collect unprecedented amounts of data on the LSS, which will enable detailed assessments of the Hubble and $S_8$ tensions in addition to shedding new light on the nature of the invisible components in the Universe.

\euclid is a space mission led by the European Space Agency (ESA) with contributions from the National Aeronautics and Space Administration (NASA), aiming to study the nature and evolution of the dark Universe. The survey uses a 1.2-m-diameter telescope and two instruments, a visible-wavelength camera, and a near-infrared camera/spectrometer, to observe billions of galaxies over more than a third of the sky in optical and near-infrared wavelengths. \euclid measures the shapes \citep{2022A&A...657A..90E, 2023A&A...671A.102E, 2023A&A...671A.101E} and redshifts \citep{2020A&A...644A..31E,2021A&A...647A.117E} of galaxies, in order to determine the weak gravitational lensing and clustering of galaxies, covering a period of cosmic history over which dark energy accelerated the expansion of the Universe. These measurements will provide detailed insights into the properties of dark energy, dark matter, and gravity by probing the expansion history of the Universe and the growth rate of structures over time \citep{2021A&A...654A.148M, 2022A&A...660A..67N,2023A&A...671A.100E}. \euclid was launched on July 1 2023 and is designed to operate for six years. The survey will provide unprecedented constraints on cosmological parameters and tests of fundamental physics, as well as a rich catalogue of legacy data that can be used for a wide range of astrophysical research. The mission data will be publicly released within two years of acquisition. \euclid is one of the most ambitious and exciting space missions in the field of cosmology and will enable a thorough validation of a broad range of cosmological models.

\euclid observations will provide precise measurements of the clustering of matter over a wide range of scales, where effects due to the late-time non-linear gravitational collapse of matter need to be taken into account. A key tool in the preparation of the cosmological analyses and the interpretation of the \euclid data is the use of cosmological {\it N}-body simulations, which can follow the non-linear evolution of matter clustering. This is a numerical technique that calculates the evolution of the matter density field under the effect of gravity across cosmic time and predicts the LSS of the Universe for a given cosmological model \citep{1974ApJ...187..425P, 1978IAUS...79..409Z, 1983MNRAS.204..891K, 1985SJSSC...6...85A, 2017ComAC...4....2P, angulo2022review}. In this method, the matter density field is sampled with discrete {\it N}-body particles, whose equations of motion are solved in the Newtonian limit in an expanding Friedmann--Lema\^{i}tre--Robertson--Walker (FLRW) Universe. These simulations enable the study of the formation and growth of cosmic structures from linear to non-linear scales and predict the distribution of matter in galaxy clusters, filaments, and voids, for a range of cosmological models and parameters \citep{2003ApJ...599...31K,2004A&A...416..853D,2010MNRAS.401..775A,Li:2011vk,2013MNRAS.436..348P,  Baldi:2014ica}, as well as initial conditions \citep{PhysRevD.77.123514}. The cosmological models beyond the standard-\lcdm paradigm are expected to have left imprints that should be detectable in the \euclid observables, such as the redshift-space power spectra of galaxies or the void-size functions.

This article is part of a series that collectively explores simulations and non-linearities beyond the $\Lambda$CDM model:
\ben
    \item Numerical methods and validation \citep{KPJC6P1}.
    \item Results from non-standard simulations (this work).
    \item Constraints on $f(R)$ models from the photometric primary probes \citep{KPJC6P4}.
    \item Cosmological constraints on non-standard cosmologies from simulated Euclid probes (D'Amico et al. in prep.).
\een
(For further details, see our companion papers.) In this work, we consistently analyse large numbers of {\it N}-body simulations over a wide range of non-standard cosmological scenarios, to generate catalogues of synthetic observables for \euclid. This analysis is achieved using a pipeline that was specifically written for that task. We calculate reconstructed density fields, halo and void catalogues, halo mass functions, dark matter, and halo power spectra in real and redshift space, as well as halo bias functions. The paper is organised as follows: in Sect.~\ref{sec:nonlcdmmodels}, we introduce the analysed non-standard models; then, in Sect.~\ref{sec:simulations}, we present an overview of the analysed cosmological {\it N}-body simulations. In Sect.~\ref{sec:analysis}, we describe the analysis pipeline and the calculated quantities. We demonstrate the imprints of the non-standard models in the computed observables in Sect.~\ref{sec:interpretation} and finally, we summarise our results in Sect.~\ref{sec:summary}.

\section{Cosmological models beyond the standard \texorpdfstring{\lcdm}{LCDM} paradigm}
\label{sec:nonlcdmmodels}

\begin{figure*}
    \centering
    \includegraphics[width=0.95\textwidth]{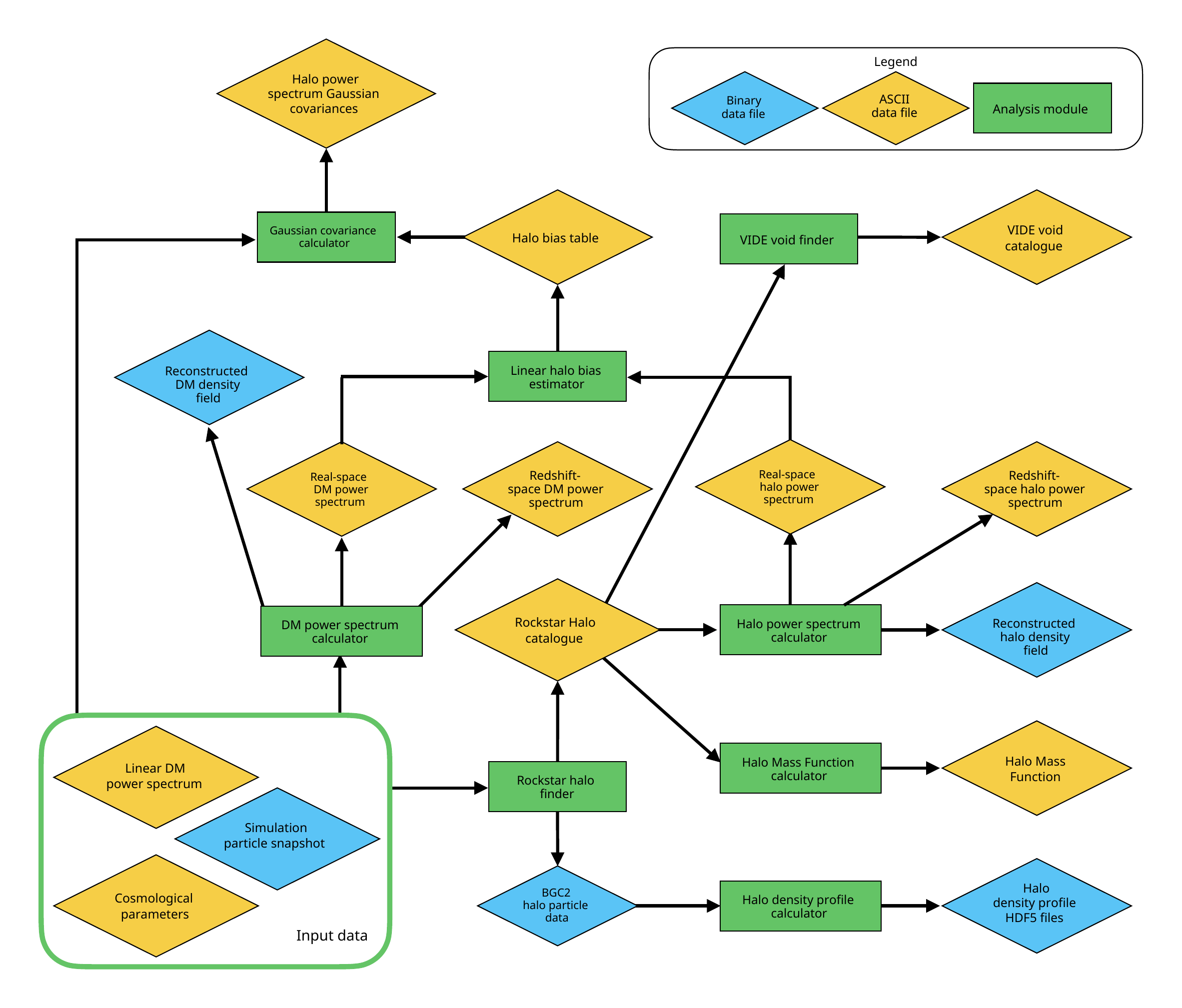}
    \caption{Flowchart summarising the main steps of the analysis pipeline. In the first step, the pipeline executes the \rockstar halo finder to generate the halo catalogues and additional BGC2 particle data containing all relevant information for the halo profile calculation. Then, the 2D and 3D halo profiles, the halo mass function, and void catalogues are calculated by the corresponding modules. The real- and redshift-space halo power spectra and the triangular shaped cloud (TSC) reconstructed halo density fields on a regular cubic grid are computed for a predefined mass bin. After this, the dark matter power spectrum calculator module computes the real- and redshift-space dark matter power spectra with the reconstructed TSC dark matter density field. By using the real-space dark matter and halo power spectra, the linear-halo-bias estimator module calculates the halo-bias table using only the linear scales. Finally, by using this halo-bias table, the linear matter power spectrum, and the cosmological parameters, the halo redshift-space power spectrum Gaussian covariances are computed. This process is repeated for all selected particle snapshots.}
    \label{fig:FlowChart}
\end{figure*}

To address the tensions and anomalies in the \lcdm model, various non-standard cosmological models have been proposed that extend or modify the standard model in different ways. Some examples of non-standard cosmological models are dark energy models, such as quintessence and phantom energy, modified-gravity theories, such as $f(R)$ gravity, and massive-neutrino models, such as sterile neutrinos and self-interacting neutrinos. These models introduce new degrees of freedom or new mechanisms that can affect the dynamics and observables of the Universe at different scales and epochs. In this section, we discuss the main features, motivations, and challenges of these non-standard cosmological models.

\subsection{Dark energy models}

\subsubsection{\texorpdfstring{\wcdm}{\textit{w}CDM}}
A simple generalisation of the cosmological constant assumes that dark energy is a fluid with a constant equation-of-state, $w \equiv p_{\de} / (\rho_{\de}\,c^2)$, where $p_{\de}$ and $\rho_{\de}$ are, respectively, the pressure and density of the fluid, and $c$ is the speed of light. To trigger an accelerated phase of cosmic expansion, the dark energy equation-of-state parameter must be $w<-1/3$. The $\Lambda$CDM model corresponds to the $w = -1$ specific case, while $w<-1$ corresponds to so-called phantom dark energy models \citep{Caldwell:2003vq}, though such values may also result from an unaccounted interaction between dark energy and dark matter \citep{2006PhRvD..73h3509D}. 

\subsubsection{Dynamical dark energy}

The dark energy equation-of-state could be a function of redshift. Chevallier, Polarski \citep{2001IJMPD..10..213C} and Linder \citep{PhysRevLett.90.091301} proposed a simple parameterisation of
\begin{equation}
w_{\de}(z) = w_0 + w_a \frac{z}{1+z} = w_0 + w_a \left(1-a\right) \, ,
\label{eq:wzCPL}
\end{equation}
where the $w_0$ parameter represents the value of the equation-of-state at the present time, and $w_a$ defines the rate of change with redshift. This model is also called the CPL parametrisation of dark energy, after the initials of the authors who proposed it. 

This dark energy parametrisation is a fitting function of a general $w_{\de}(z)$ around $z=0$, assuming that $w(z)$ is smooth and slowly changing with the scale factor. As a consequence, this model can closely follow the expansion history of a wide range of other models with $w_{\de}(z)$ at late times. Despite its simple form, it shows a wide range of interesting properties \citep{2008GReGr..40..329L, 2008PhRvD..78b3526L}.  The cosmological constant corresponds to  $w_0 = -1$ and $w_a = 0$ in the CPL parametrisation.

\subsubsection{K-essence}

The k-essence model is characterised by an action for the scalar field of the following form:
\begin{equation}
S=\int \diff ^4 x\;\sqrt{-g}\;p(\phi,X)\;,
\label{eq:action_kessence}
\end{equation}
where $X=(1/2)g^{\mu\nu}\nabla_{\mu}\phi\nabla_{\nu}\phi$.  The energy density of the scalar field is given by 
\begin{equation}
\energydensity_{\phi}=2X\;\frac{\diff p}{\diff X}-p\;,
\end{equation}
and the pressure is $p_{\phi}=p(\phi,X)$. This pressure gives an effective fluid equation-of-state parameter as
\begin{equation}
w_{\phi}=\frac{p_{\phi}}{\energydensity_{\phi}}=-\frac{p}{p-2X p,_{X}}\;,
\end{equation}
where the subscript $,_{X}$ indicates a derivative with respect to $X$, and a dimensionless speed-of-sound parameter for the k-essence fluctuations as 
\begin{equation}
c_{\rm s}^2 =\frac{p,_{X}}{p,_X+2Xp,_{XX}} \;.
\end{equation}
The k-essence field satisfies the continuity equation
\begin{equation}
\dot{\energydensity}_{\phi}=-3H\,(\energydensity_{\phi}+p_{\phi})\;,
\end{equation}
which results in the scalar equation of motion
\begin{equation}
G^{\mu\nu}\nabla_{\mu}\nabla_{\nu}\phi+2X\frac{\partial^2p}{\partial X \partial \phi}-\frac{\partial p}{\partial \phi}=0\;,
\end{equation}
where
\begin{equation}
G^{\mu\nu}=\frac{\partial p}{\partial X}g^{\mu\nu}+\frac{\partial^2 p}{\partial X^2}\nabla^{\mu}\phi\nabla^{\nu}\phi\;.
\end{equation}
The k-essence was first proposed by \cite{Armendariz-Picon:2000nqq,Armendariz-Picon:2000ulo}, who showed that there exist tracking attractor solutions to the equation of motion during the radiation and matter-dominated eras of the Universe, and that with a suitably chosen $p$, the scalar can have an appropriate equation of state that allows it to act as dark energy for the background accelerated expansion. In addition, whenever the kinetic terms for the scalar field are not linear in $X$, the speed of sound of the fluctuations differs from unity, allowing for the clustering of the dark energy field at sub-horizon scales, which should be modelled at the perturbations level.

\subsubsection{Interacting dark energy}
\label{sec:iDE}

In the interacting dark energy (IDE) models \citep{Amendola:1999er,Farrar:2003uw,Baldi:2008ay}, dark energy and cold dark matter are allowed to interact through an exchange of energy-momentum in order to keep the total stress-energy tensor, $T_{\mu \nu}$, conserved:
\begin{equation}
    \nabla _{\mu }T^{(c)\mu }_{\nu} = C_{\nu}(\phi ) = - \nabla _{\mu }T^{(\phi )\mu }_{\nu} \,,
\end{equation}
where $C_{\nu }(\phi )$ is a conformal coupling function expressed in the form:
\begin{equation}
    C_{\nu }(\phi ) = \kappa \, \beta(\phi)\,\energydensity _{\rm c}\nabla _{\nu }\phi \,,
\end{equation}
where $\kappa \equiv \frac{8\pi G_{\rm N}}{c^2}$, $G_{\rm N}$ is Newton's gravitational constant, $\energydensity _{\rm c}$ is the cold dark matter energy density in the IDE model \footnote{Note that we choose $u_c$ here for our IDE model to better distinguish it from $\rho_{CDM}$ used right afterwards to describe the \lcdm background evolution.}, and $\beta (\phi )$ is a coupling function.
The dark energy scalar field, $\phi$, has an intrinsic energy density and pressure given by
\begin{eqnarray}
\label{phidensity}
    \energydensity _{\phi } &=& \frac{1}{2}\,g^{\mu \nu }\,\partial _{\mu }\phi \,\partial _{\nu }\phi + V(\phi )\,,\\
    \label{phipressure}
    p_{\phi } &=& \frac{1}{2}\,g^{\mu \nu }\,\partial _{\mu }\phi \,\partial _{\nu }\phi - V(\phi )\,,
\end{eqnarray}
where $V(\phi )$ is a self-interaction potential. The conservation equations then translate into the following set of background-dynamic equations under the assumption of a constant coupling function $\beta (\phi )= \beta $:
\begin{eqnarray}
\label{kleingordon}
\ddot{\phi } + 3H\dot{\phi } + \frac{\diff V}{\diff\phi }&=&\kappa \, \beta \, \energydensity _{\rm c} \,,\\
\label{continuity}
\dot{\energydensity }_{\rm c} + 3H\energydensity _{\rm c}&=&-\kappa \,\beta \, \energydensity _{\rm c} \,,
\end{eqnarray}
In the standard approach, a theoretically motivated analytical form for the self-interaction potential function, $V(\phi )$, is chosen. However, the simulations that are considered in the present work implement the alternative approach proposed by \citet{Barros_etal_2019} that consists of imposing a standard $\Lambda $CDM background expansion history by setting 
\begin{equation}
\label{constraint}
 H^{2}=H^{2}_{\Lambda {\rm CDM}} \,,  
\end{equation}
where $H_{\Lambda {\rm CDM}}$ is the standard Hubble function defined by
\begin{equation}
    H^{2}_{\Lambda {\rm CDM}} = \frac{8 \pi G_{\rm N}}{3}(\rho _{\rm r} + \rho _{\rm b} + \rho _{\rm CDM} + \rho _{\Lambda })\,,
\end{equation}
where $\rho_{\rm r}$, $\rho_{\rm b}$, $\rho_{\rm CDM}$, and $\rho_{\Lambda }$ are the mass densities of the radiation, baryon, CDM, and $\Lambda$ components of the background \lcdm model. This will determine an effective potential, $V(\phi )$, according to the resulting evolution of the scalar field, $\phi $. Taking the time derivative of Eq.~(\ref{constraint}) and using the continuity Eqs. (\ref{kleingordon} \& \ref{continuity}), one gets the scalar-field energy density and pressure as
\begin{eqnarray}
\label{newenergy}
\energydensity _{\phi } &=& \rho _{\rm CDM}\,c^2 + \rho _{\Lambda }\,c^2 - \energydensity _{\rm c}\,, \\
\label{newpressure}
p_{\phi }&=&p_{\Lambda }=-\energydensity _{\Lambda} \, ,
\end{eqnarray}
which can be combined with Eqs.~(\ref{phidensity} \& \ref{phipressure}) to obtain the dynamics of the scalar field:
\begin{equation}
\label{mainrelation}
    \dot{\phi }^{2} = \rho _{\rm CDM}\,c^2 - \energydensity _{\phi }\,.
\end{equation}
The scalar-field potential, $V(\phi )$, can then be reconstructed using Eqs.~(\ref{newenergy} \& \ref{newpressure}) as:
\begin{equation}
\label{potential}
    V(\phi ) = \frac{1}{2}\dot{\phi }^{2} + \rho _{\Lambda }\,c^2\,,
\end{equation}
and taking the time derivative of Eq.~(\ref{mainrelation}), one can derive the scalar-field equation of motion
\begin{equation}
\label{field}
    2\ddot{\phi } + 3H\dot{\phi } -\kappa \beta \energydensity _{c} = 0 \,,
\end{equation}
which can be numerically solved for the dynamical evolution of the system. With this choice, the $\beta$ coupling remains the only free parameter of this model.
Observational constraints on the model were computed in \cite{Barros:2022bdv}, which found that the model can alleviate the $\sigma_8$ tension, but that CMB prefers the $\Lambda$CDM limit. In particular, they find that the CMB constrains $|\beta| \lesssim 0.02$, RSD constraints $|\beta| \lesssim 0.10$, while weak lensing data from the Kilo-Degree Survey actually prefers a non-zero value $|\beta| \sim 0.1$.

\subsection{Modified gravity models}

\subsubsection{nDGP gravity}
The Dvali--Gabadadze--Porrati (DGP) model \citep{Dvali:2000hr} assumes that our Universe is described by a five-dimensional bulk, while the visible matter component is confined to the four-dimensional brane described by the Minkowski metric, $\gamma$. 
This model's action is 
\begin{equation}
S = \frac{c^4}{16\pi G_5}\int_{\cal M} {\rm d}^5 x \sqrt{-\gamma}\,R_5
+ \int_{\partial {\cal M}} {\rm d}^4 x \,\sqrt{-g}\, 
\left(\frac{c^4}{16\pi G_{\rm N}}\,R + {\cal L}_{\rm m} \right)\,,
\end{equation}
where $G_5$ and $G_{\rm N}$ are the five- and four-dimensional Newton's constants, respectively, and ${\cal L}_{\rm m}$ is the matter Lagrangian. At small scales, four-dimensional gravity is recovered due to an intrinsic Einstein–Hilbert term sourced by brane curvature causing a gravitational force that scales as $r^{-2}$, while, at large scales, the gravity behaves as a five-dimensional force. The transition between the five-dimensional modifications and the four-dimensional gravity is given by the cross-over scale, $r_{\rm c} = G_5/(2 G_{\rm N})$, from which we construct the dimensionless parameter $\Omega_{\rm rc} \equiv c^2/(4r_{\rm c}^2H_0^2)$. The modified Friedmann equation on the brane \citep{Deffayet:2000uy} becomes
\begin{equation}\label{eqn:HDGP}
 H^2 = \pm\, c \frac{H}{r_{\rm c}} + \frac{8 \pi G_{\rm N}}{3} \bar{\rho}\,.
\end{equation}
The model we investigate in this paper is the normal branch with the $-$ sign \citep{Bowcock:2000cq} characterised by a \lcdm background achieved by introducing an additional dark energy contribution with an appropriate equation-of-state \citep{Schmidt:2009sv},
\begin{equation}
\rho_{\de}(a) = \rho_{\rm cr,0} \, \left( \ol + 2 \sqrt{\Omega_{\rm rc}} \sqrt{\ol + \om \, a^{-3}} \, \right)\,,
\end{equation}
where $\rho_{\rm cr,0}$ is the critical density.
The observational constraints on the model require the cross-over scale, $r_{\rm c}$, to be larger than the size of the horizon $H_0^{-1}$ today. For example, Solar System constraints require $r_cH_0 \gtrsim 1.6$ \citep{Battat:2008bu}, and galaxy clustering in the BOSS survey constraints $r_cH_0 \gtrsim 4.5$ \citep{Piga:2022mge}.

\subsubsection{\texorpdfstring{$f(R)$}{\textit{f}(\textit{R})} gravity}
The $f(R)$ theory of gravity \citep{Buchdahl:1970ynr} is characterised by the following action:
\begin{align}
 S = \frac{c^4}{16\pi G_{\rm N}} \int{{\rm d}^4 x \,\sqrt{-g}\, \left[\,R+f(R)\,\right]} \,, \label{eq:EHaction}
\end{align}
where $g_{\mu\nu}$ is the metric tensor and $f(R)$ is a functional form of the Ricci scalar, $R$.
Here, we consider the Hu--Sawicki model \citep{Hu:2007nk}  with $n=1$,  where in the limit of $f_R=\diff f/\diff R\ll1$ we have
\begin{equation}
f(R) = - 6 \ol\,\frac{H_0^2}{c^2} + |f_{R0}|\,\frac{\bar{R}_0^2}{R}\,,\label{eq:fR}
\end{equation}
where $f_{R0}$ is the free parameter of the model,
$\bar R_0$ is the Ricci scalar evaluated at background at present time, $H_0$ is the Hubble constant, and $\ol$ is the energy-density parameter of the cosmological constant.
$|f_{R0}|$ characterises the magnitude of the deviation from \lcdm, with smaller values corresponding to weaker departures from general relativity until we recover \lcdm\ in the limit of $f_{R0}\rightarrow0$, but for the small $|f_{R0}|$ values still allowed by observations, the background expansion history approximates that of \lcdm\ and
\begin{equation}
\bar{R}_0 = 3 \om \frac{H_0^2}{c^2} \left(1+ 4 \frac{\ol}{\om} \right)\,,\label{eq:R}
\end{equation}
with the matter energy density parameter $\om=1-\ol$. However, though the background expansion could mimic that of a cosmological-constant model, it still differs at the level of cosmological perturbations where the growth of structure is driven by a modification of gravity following the above adopted model of $f(R)$.

The observational constraints on the model parameter $|f_{R0}|$ vary from $|f_{R0}| \lesssim 10^{-6}$ in the Solar System, $|f_{R0}| \lesssim 10^{-8}$ from galaxy scales \citep{Burrage:2023eol} to $|f_{R0}| \lesssim 10^{-6}$--$10^{-4}$ from various cosmological probes \citep[see, e.g., Fig. 28 in][for a summary]{Koyama:2015vza}. The parameter values of the simulations presented in this paper are similar to the current cosmological constraints.

\subsection{Massive relativistic neutrinos}
Neutrinos are mainly characterised by two properties: their mass, $M_{\nu}$, and the number of neutrino species, $\neff$. More in general, $\neff$ parametrises the contribution of relativistic species to the background density of radiation, $\rho_{\rm r}$, as
\begin{equation}
 \rho_{\rm r} = \left[1 + \frac{7}{8}\left(\frac{4}{11}\right)^{4/3}\neff\right]\rho_{\gamma}\,,
\end{equation}
where $\rho_{\gamma}$ is the photon background density. In the standard model, $N_{\rm eff}$ is expected to be $\sim$ 3.045 \citep{Cielo:2023bqp} for three families of active neutrinos that thermalised in the early Universe and decoupled well before electron-positron annihilation. The calculation of $N_{\rm eff}$ involves the complete treatment of neutrino decoupling, which incorporates non-instantaneous decoupling. A deviation from the fiducial value serves to account for the presence of non-standard neutrino features, or additional relativistic relics contributing to the energy budget \citep{Mangano:2001iu}. Here we focus on standard neutrino families only.\\

In addition, oscillation experiments \citep{Maltoni:2004ei,Kajita:2016cak} showed that at least two neutrinos are massive by measuring two squared-mass differences.
It can be shown that the minimum value of the neutrino mass sum is either $0.06\,{\rm eV}$ in the normal or $0.10\,{\rm eV}$ in the inverted hierarchy. This value can be well constrained through cosmological observations since neutrinos are known to impact the expansion history and suppress the clustering of cold dark matter, which can be observed in the large-scale distribution of galaxies \citep{Sakr:2022ans}.
Neutrinos with mass $\lesssim 0.6$ eV become
non-relativistic after the epoch of recombination probed by the CMB,
and this mechanism allows massive neutrinos to alter the
matter-radiation equality for a fixed $\om h^2$
\citep{lesgourgues2006}. Massive neutrinos act as non-relativistic
particles on scales $k>k_{\rm nr}=0.018(m_\nu/1{\rm
  eV})^{1/2}\om^{1/2}$ \hMpc, where $k_{\rm nr}$ is the
wavenumber corresponding to the Hubble horizon size at the epoch
$z_{\rm nr}$ when the given neutrino species becomes non-relativistic following $1+z_\mathrm{nr} \simeq 1900 \left( \frac{m_\nu}{\rm{1 eV}} \right)$,
$\om$ is the matter density parameter, and $h=H_0/100\, {\rm km\,
  s^{-1} Mpc^{-1}}$. The large velocity dispersion of non-relativistic
neutrinos suppresses the formation of neutrino perturbations in a way
that depends on $m_\nu$ and redshift $z$, leaving an imprint on the
matter power spectrum at scales $k>k_{\rm fs}(z)$, with 
\begin{equation}
k_{\rm fs}=\frac{0.82 H(z)}{H_0(1+z)^2} \left( \frac{m_\nu}{\rm{1 eV}} \right)\homopc\,,
\end{equation}
where neutrinos cannot cluster and do not contribute to the gravitational potential wells produced by cold dark matter and baryons~\citep{takada2006,lesgourgues2006}. This modifies the shape of the matter power spectrum and the correlation function on these scales.

\subsection{Primordial non-Gaussianities}

The simplest inflation models predict that primordial curvature perturbations follow a distribution that is close to Gaussian \citep{Maldacena_2003,Creminelli_2004}. However, there are many alternative inflation models that predict certain amounts of primordial non-Gaussianity (PNG). One of the simplest cases is that of the so-called local primordial non-Gaussianities \citep{Salopek_1990,Komatsu_2001}. For this case, the primordial potential $\phi$ is given by 
\begin{equation}
    \phi(\textbf{x}) = \phi_G(\textbf{x}) + f_{\rm NL}^{\rm local} (\phi^2(\textbf{x}) - \langle \phi^2(\textbf{x}) \rangle), 
\end{equation}
where $\phi_G(\textbf{x})$ is the Gaussian potential, while $\phi$ is the non-Gaussian potential. $f_{\rm NL}^{\rm local}$ measures the level of deviations from Gaussianity.

The perturbations in the primordial potential produce perturbations in the density field and they are related through Poisson's equation. Therefore, in Fourier space, the density field is given by
\begin{equation}
\delta(k,z)\; = \;\alpha(k,z)\;\phi(k,z),
\end{equation}
where
\begin{equation}
    \alpha(k,z) = \frac{2 D(z)}{3 \om} \frac{c^2}{H_0^2} \frac{g(0)}{g(z_{\rm rad})} k^2 T(k),
\end{equation}
$T(k)$ is the transfer function normalised at $T(k\rightarrow0)=1$, and  $D(z)$ is the growth factor normalised at $D(z=0) =1$. The factor $g(0)/g(z_{\rm rad})$, where $g(z)= (1+z)D(z)$, takes into account the difference between our normalisation of $D(z)$ and the early-time normalisation where $D(z) \propto 1/(1+z)$ during matter-domination. This factor is $ \frac{g(z_{\rm rad})}{g(0)} \sim 1.3$, with a small dependency on the cosmology.

This type of non-Gaussianity characteristically affects the clustering of biased tracers, inducing a scale-dependent bias \citep{PhysRevD.77.123514, Slosar_2008, Matarrese_2008}. To linear order, the power spectrum of galaxies can be given as
\begin{equation}
    P_{\rm t,t}(k,z) = \left[b_1 + \frac{b_\phi f_{\rm NL}^{\rm local}}{\alpha(k,z)}\right]^2 P_{\rm m,m}(k,z), 
\end{equation}
where $P_{\rm t,t}(k,z)$ is the power spectrum of the tracer, $P_{\rm m,m}(k,z)$ is the power spectrum of the matter, $b_1$ is the linear bias, and $b_\phi$ is the response of the tracer to the presence of the local-PNG. Now, $P_{\rm t,t}(k,z)$ has a dependency with $k$ which scales as $k^{-2}$ at leading order due to the $\alpha(k,z)$ term.
The $b_\phi$ is usually parametrised as
\begin{equation}
    b_\phi = 2 \delta_c (b_1 -p).
\end{equation}

Although it is possible to make a theoretical prediction for $p$ (by assuming a universal mass function, $p=1$, \citealp{PhysRevD.77.123514}), several studies using numerical simulations have shown that the prediction may be different depending on the type of galaxy or tracer under consideration \citep{Slosar_2008, Desjacques_2009, Hamaus_2011,Biagetti_2017, Barreira_2020,Adame24}. 

\section{Simulations}
\label{sec:simulations}

\begin{figure}
    \centering
    \includegraphics[width=0.49\textwidth]{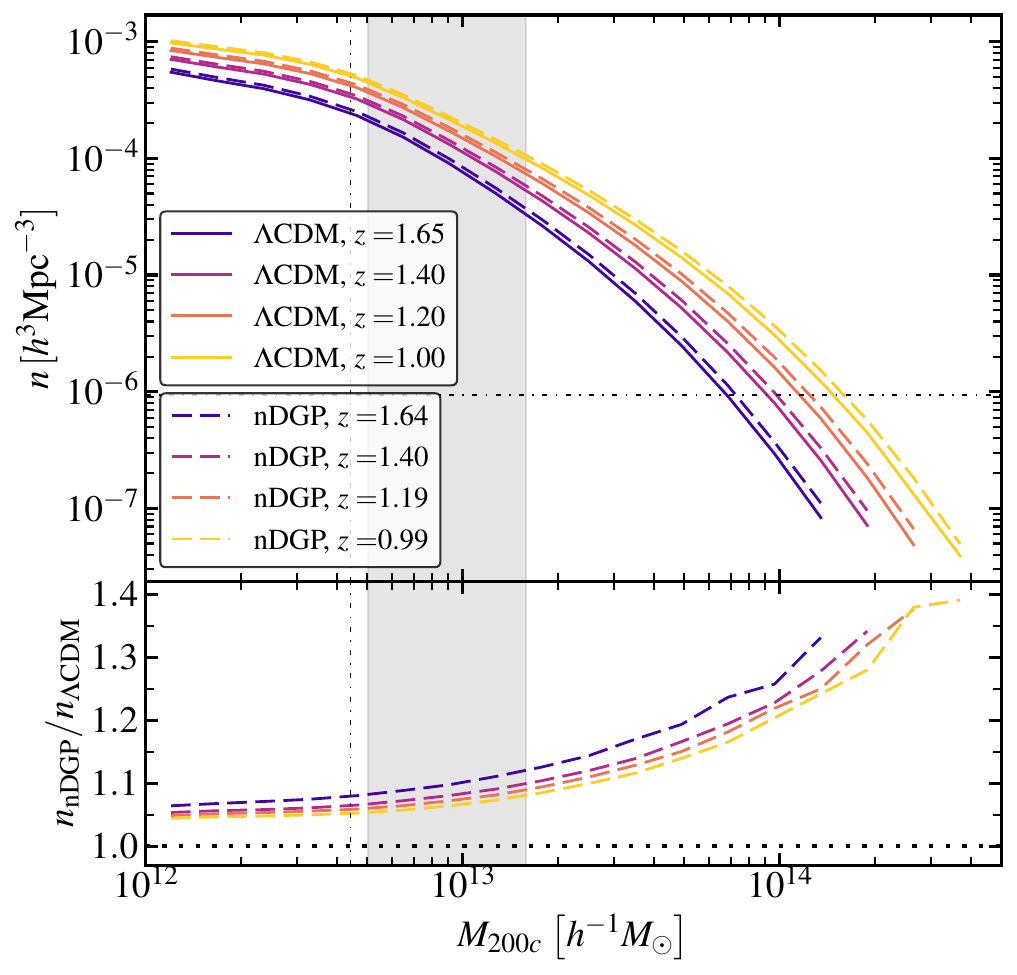}
    \caption{Calculated halo mass function of the \ELEPHANT simulation suite. The vertical and horizontal dash-dotted lines indicate the mass relative to haloes with 50 particles and the number density relative to a $1\%$ shot-noise error, respectively. The shaded region highlights the mass bin used to calculate the halo power spectra shown in Fig.~\ref{fig:ElephantPkComparison}.} 
    \label{fig:ElephantHMFComparison}
\end{figure}

\begin{figure*}
    \centering
    \includegraphics[width=0.99\textwidth]{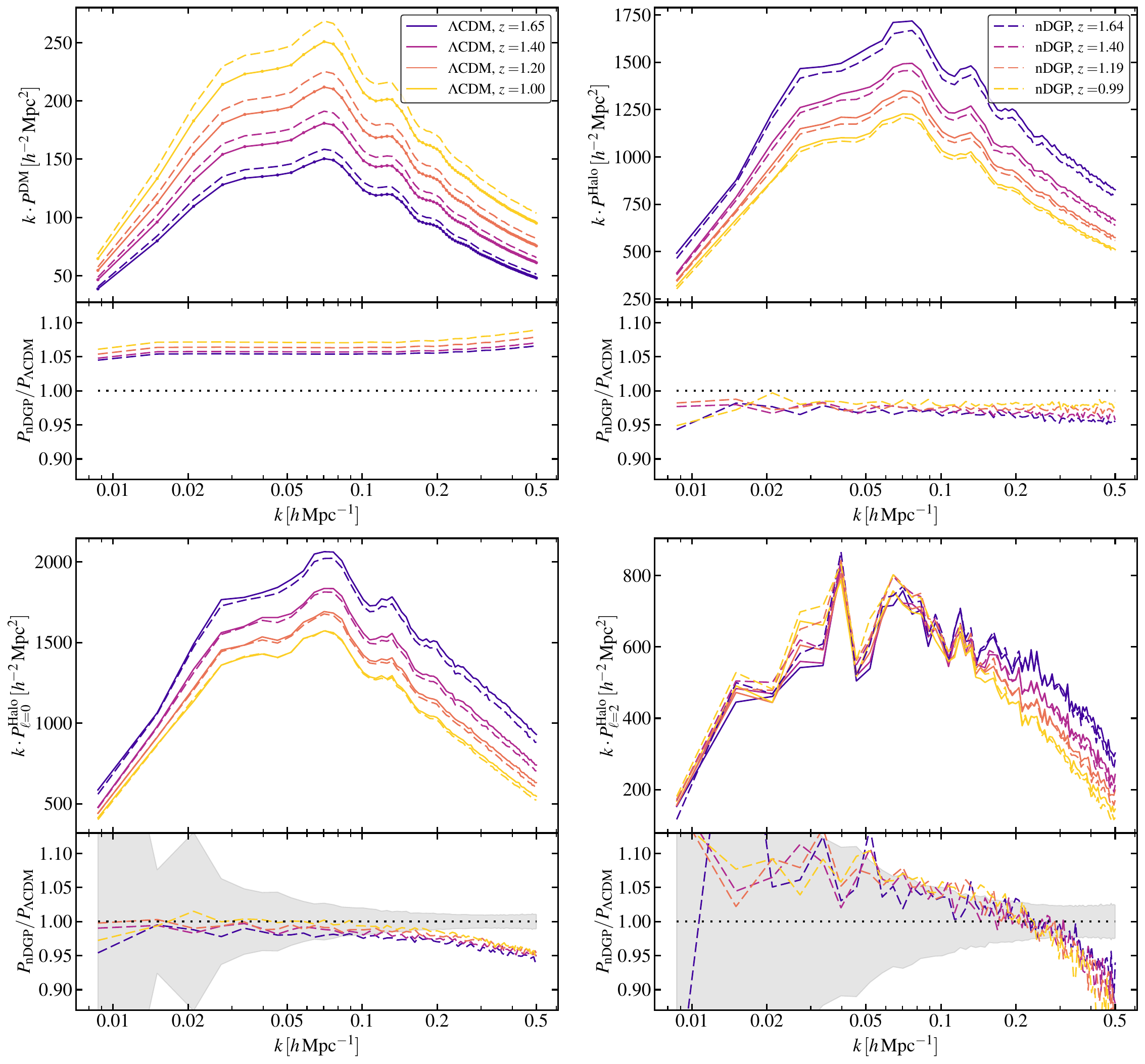}
    \caption{Calculated matter and halo power spectra of the \ELEPHANT simulation suite in the mass bin $10^{12.7}\msoh<M_{\rm halo}<10^{13.2}\msoh$. The solid lines represent the reference \lcdm simulations, while the dashed lines the results of the nDGP simulations. \textit{Top left:} Real-space power spectra for dark matter. The dots above the solid lines highlight the locations where the power spectrum is estimated. \textit{Top right:} Real-space power spectra for haloes. \textit{Bottom left:} Monopole of the halo power spectrum in redshift space. The shaded region shows the calculated variance of the monopole of the \lcdm halo power spectrum at redshift $z=1$. \textit{Bottom right:} Quadrupole of the halo power spectrum in redshift space. The shaded region represents the calculated variance of the quadrupole of the \lcdm halo power spectrum at redshift $z=1$.} 
    \label{fig:ElephantPkComparison}
\end{figure*}

This section summarises the simulations used for this project and gives a very brief description of each setup. The analysed simulations followed the evolution of the matter field with discrete $N$-body method in the models described in Sect.~\ref{sec:nonlcdmmodels}. Baryonic and hydrodynamical effects are neglected in this paper. For a comprehensive description of each of the simulation suites, we refer the reader to the main references given in Table \ref{tab:sims} along with the volumes, resolutions, initial redshifts, and the used order of the Lagrangian perturbation theory (LPT) during the initial-condition generation.
\begin{table*}[]
    \centering
     \caption{Overview of the simulation suites analysed for this project.}
    \begin{tabular}{l l r r r c c c c}
    \toprule
    Name/Reference & Code & $N_{\rm sim}$ & $L$ [$\gpcoh$] & $N_\mathrm{DM}$ & $m_\mathrm{DM}$ [$\msoh$] & $z_{\rm init}$ & LPT order & Model \\
    \midrule
    \COMPLEMENTARY  & \multirow{2}{*}{\GIZMO} & $2$ & $1.5\phantom{00}$ & $2160^3$ & $2.89\times10^{10}$ & \multirow{2}{*}{$127$} & \multirow{2}{*}{1LPT} & \lcdm  \\
    1 & & $2$ & $1.5\phantom{00}$ & $2160^3$ & $2.88\times10^{10}$ & & & \wcdm \\
    \midrule
    \multirow{2}{0.17\linewidth}{\DEMNUni}  & \PGadgetThree & $100$ & $1.0\phantom{00}$ & $1024^3$ & $8\times10^{10}$ & \multirow{3}{*}{$99$} & \multirow{3}{*}{1LPT} & \lcdm$ +m_{\nu}$  \\
     & \PGadgetThree & $15$ & $2.0\phantom{00}$ & $2048^3$ & $8\times10^{10}$ &  & & ${\rm CPL} +m_{\nu}$   \\
    2 & \PGadgetThree & $2$ & $0.5\phantom{00}$ & $2048^3$ & $1.25\times10^{9}$ & & & \lcdm$ +m_{\nu}$ \\
    \midrule
    \RAYGAL & \multirow{2}{*}{\RAMSES} & 1 & $2.625$ & $4096^3$ & $1.88\times10^{10}$ & \multirow{2}{*}{49} & \multirow{2}{*}{2LPT} & \lcdm  \\
    3 & &  1 & $2.625$ &  $4096^3$ & $2.0\times10^{10}$ & &  & \wcdm \\
    \midrule
    \ELEPHANT  & \multirow{2}{*}{\ECOSMOG} & $11$ & $1.024$ & $1024^3$ & $8.85\times10^{10}$ & \multirow{2}{*}{$49$} & \multirow{2}{*}{2LPT} & \lcdm \\
    4 & & $11$ & $1.024$ & $1024^3$ & $8.85\times10^{10}$ & & & nDGP \\
    \midrule
    \ColaHiRes  & \multirow{2}{*}{\COLA} & $5$ & $1.024$ & $2048^3$ & $1.07\times10^{10}$ & \multirow{2}{*}{$127$} & \multirow{2}{*}{2LPT} & nDGP  \\
    5 & & $2$ & $1.024$ & $2048^3$ & $1.07\times10^{10}$ & & & \lcdm  \\
    \midrule
    \DUSTGRAIN  & \multirow{2}{*}{\MGGADGET} & $3$ & $2.0\phantom{00}$ & $2048^3$ & $8.27\times10^{10}$ & \multirow{2}{*}{$99$} & \multirow{2}{*}{1LPT} & nDGP \\
    6 &  & $11$ & $0.75\phantom{0}$ & $768^3$ & $8.1\times10^{10}$ & & & $f(R)+m_{\nu}$   \\
    \midrule
    \CIDER  & \multirow{2}{*}{\CGADGET} & \multirow{2}{*}{$4$} & \multirow{2}{*}{$1.0\phantom{00}$} & \multirow{2}{*}{$1024^3$} & \multirow{2}{*}{$8.1\times10^{10}$} & \multirow{2}{*}{$99$} & \multirow{2}{*}{2LPT} & \multirow{2}{*}{cDE} \\
    7 & & & & & & & &  \\
    \midrule
    \DAKARONEANDTWO & \multirow{2}{*}{\CGADGET} & \multirow{2}{*}{$5$} & \multirow{2}{*}{$1.0\phantom{00}$} & \multirow{2}{*}{$1024^3$} & \multirow{2}{*}{$8.1\times10^{10}$} & \multirow{2}{*}{$99$} & \multirow{2}{*}{1LPT} & \multirow{2}{*}{DS}  \\
    8 & & & & & & & &  \\
    \midrule
    \ClusteringDE & \multirow{2}{*}{\KEVOLUTION} & 1 & $2.0\phantom{00}$ & $2400^3$ & $4.2\times 10^{10}$ & \multirow{2}{*}{$100$} & \multirow{2}{*}{1LPT} & $w_0c_{\rm s}^2$CDM   \\
    9, 10 & &  1 & $2.0\phantom{00}$ &  $1200^3$ & $3.3\times 10^{11}$ & & & $w_0c_{\rm s}^2$CDM  \\
    \midrule
    \FORGE & \multirow{2}{*}{\MGAREPO} & $100$ & $0.5\phantom{00}$ & $1024^3$ & $\sim10^{10}$ & \multirow{2}{*}{$127$} & \multirow{2}{*}{2LPT} & \lcdm{}  \\
    11 &  & $98$ & $0.5\phantom{00}$ & $1024^3$ & $\sim10^{10}$ & & & $f(R)$ \\
    \midrule
    \BRIDGE & \multirow{2}{*}{\MGAREPO} & \multirow{2}{*}{$98$} & \multirow{2}{*}{$0.5\phantom{00}$} & \multirow{2}{*}{$1024^3$} & \multirow{2}{*}{$\sim10^{10}$} & \multirow{2}{*}{$127$} & \multirow{2}{*}{2LPT} & \multirow{2}{*}{nDGP} \\
    12 & & & & & & & &  \\
    \midrule
    \PNGUNITsim & \multirow{2}{*}{\GADGETTWO} & \multirow{2}{*}{1} & \multirow{2}{*}{$1.0\phantom{00}$} & \multirow{2}{*}{$4096^3$} & \multirow{2}{*}{$1.2\times10^9$} & \multirow{2}{*}{$99$} & \multirow{2}{*}{2LPT} & \multirow{2}{*}{PNG}  \\
    13 & & & & & & & &  \\
    \bottomrule
    \end{tabular}
    \tablefoot{All simulations were dark matter only, except the \CIDER and \DAKARTWO suites, which used a two-component collisionless approximation to follow the baryonic and dark matter components separately. Hydrodynamic simulations were not analysed in this project.}
    \tablebib{(1)~\citet{2023AA...672A..59R}; (2) \citet{DEMNUni_simulations}; (3) \citet{rasera2022raygal}; (4) \citet{2018MNRAS.476.3195C}; (5) \citet{Fiorini:2023fjl}; (6) \citet{Giocoli:2018gqh}; (7) \citet{Baldi:2022uwb}; (8) \citet{Baldi:2016zom}; (9) \citet{Hassani_2019}; (10) \citet{Hassani:2019wed}; (11) \citet{2021arXiv210904984A}; (12) \citet{Harnois-Deraps:2022bie}; (13) \citet{Adame24}.}
    \label{tab:sims}
\end{table*}

\subsection{The \COMPLEMENTARY simulations}

The \COMPLEMENTARY simulation series is a set of 4 cosmological {\it N}-body simulations in \wcdm and \lcdm cosmologies.
This suite used the complementary-simulation method \citep{2023AA...672A..59R}, which is a novel technique in which cosmological {\it N}-body simulations are run in phase-shifted matching pairs. One simulation starts from a regular random Gaussian initial condition, while the second simulation has modified initial amplitudes of the Fourier modes to ensure that the average power spectrum of the pair is equal to the cosmic mean power spectrum from linear theory at the initial time. The average statistical properties of a pair of such simulations have greatly suppressed variance. In this paper, we have analysed two complementary pairs using \lcdm and \wcdm cosmologies.
The \lcdm simulation pair used the best-fit Planck2018 \citep{2020AA...641A...6P} cosmological parameters:
$\om=1-\ol=0.3111$, $\ob=0.04897$, $H_0=67.66\kmsmpc$, $n_{\rm s}=0.9665$, and $\sigma_8=0.8102$. The \wcdm pair had the following parameters: $w_0=-1.04$, $\om=0.3096$, $\ob=0.04899$, $\ode=0.6904$, $H_0=67.66\kmsmpc$, $n_{\rm s}=0.9331$, and $\sigma_8=0.8438$.
The cosmological simulations of this series were run using the cosmological {\it N}-body code \GIZMO \citep{2015MNRAS.450...53H}.
All simulations in the series contained $2160^3$ dark matter particles in a $(1.5\gpcoh)^3$ volume, with $\varepsilon=13.8\kpcoh$ softening length.
The initial conditions (ICs) were generated by a modified version of the \NGENIC code \citep{2015ascl.soft02003S} by using the Zeldovich approximation
and initial linear power spectra from the Boltzmann code \CAMB \citep{2011ascl.soft02026L}.
The simulations started from redshift $z_{\rm init}=127$, with a total of 48 output times. In this project, 31 particle snapshots were analysed in the $0.5\leq z \leq 2.0$ redshift range for each simulation.

\subsection{The \DEMNUni simulation suite}

The `Dark Energy and Massive Neutrino Universe' (\DEMNUni) simulations~\citep{DEMNUni_simulations,Parimbelli2022} have been produced with the aim of investigating the LSS in the presence of massive neutrinos and dynamical dark energy, and they were conceived for the non-linear analysis and modelling of different probes, including dark matter, halo, and galaxy clustering \citep[see][]{DEMNUni1,Zennaro2018,Parimbelli2022, Gouyou_Beauchamps_2023}, weak lensing, CMB lensing, Sunyaev-Zeldovich, and integrated Sachs-Wolfe (ISW) effects \citep{Roncarelli2015,DEMNUni_simulations}, cosmic void statistics \citep{Kreisch2019}, and cross-correlations among these probes \citep{Cuozzo2022}.
The \DEMNUni simulations were run using the tree particle mesh-smoothed particle hydrodynamics (TreePM-SPH) code \PGadgetThree \citep{2005MNRAS.364.1105S}, specifically modified as in \cite{Viel_2010} to account for the presence of massive neutrinos. This modified version of \PGadgetThree follows the evolution of CDM and neutrino particles, treating them as two distinct collisionless components.
The reference cosmological parameters were chosen to be close to the baseline Planck 2013 cosmology \citep{Planck2013_XVI}:
$\ob=0.05$, $\om=0.32$, $H_0=67.0\kmsmpc$, $n_{\rm s}=0.96$, and $A_{\rm s} =2.127 \times 10^{-9}$.
Given these values, the reference (i.e. the massless neutrino case) CDM-particle mass resolution is $m^{\rm p}_{\rm CDM} = 8.27\times 10^{10} \, \msoh$, which is decreased according to the mass of neutrino particles, in order to keep the same $\om$ among all the \DEMNUni simulations. In fact, massive neutrinos are assumed to come as a particle component in a three-mass-degenerate scenario, therefore, to keep $\om$ fixed, an increase in the massive neutrino density fraction yields a decrease in the CDM density fraction.
The \DEMNUni simulations balance mass resolution and volume to include perturbations at both large and small scales. The simulations are characterised by a softening length of $\varepsilon=20\kpcoh$, a comoving volume of $8 \gpcohcube$ filled with $2048^3$ dark matter particles and, when present, $2048^3$ neutrino particles. The simulations are initialised at $z_{\rm init}=99$ with Zeldovich initial conditions. The initial power spectrum is rescaled to the initial redshift via the rescaling method developed in~\cite{zennaro_2017}. Initial conditions are then generated with a modified version of the \NGENIC software, assuming Rayleigh random amplitudes and uniform random phases.

\subsection{The \RAYGAL simulations}
The \RAYGAL simulations \citep{breton2019imprints,rasera2022raygal} are a set of two dark-matter only simulations in \wcdm and \lcdm cosmologies. The simulations were performed with the adaptive-mesh refinement (AMR) {\it N}-body code \RAMSES \citep{teyssier2002cosmological, guillet2011simple}. 
These simulations have a box size of $2625\mpcoh$  for $4096^3$ particles, which results in a smoothing scale of $5 \kpcoh$ at the maximum refinement level.
Both simulations share the parameters $H_0= 72.0\kmsmpc$, $n_{\rm s} = 0.963$, $\ob = 0.04356$ and $\Omega_{\rm r} = 8.076\times 10^{-5}$. The flat \lcdm simulation has a WMAP7 cosmology \citep{komatsu2011wmap7}: $\om = 0.25733$, and $\sigma_8 = 0.80101$, while the flat \wcdm simulation is consistent at the $1 \sigma$-level with a WMAP7 cosmology with $\om = 0.27508$, $\sigma_8 =0.85205$, and $w = -1.2$.
In both cases, Gaussian initial conditions are generated using a modified version of the code \MPGRAFIC \citep{prunet2008initial} with the displacement field computed using second-order Lagrangian perturbation theory (2LPT) to minimise the effect of transients \citep{2006MNRAS.373..369C}. The initial redshift has been set to $z_{\rm init} \sim 46$ such as to ensure that the maximum displacement is of the order of one coarse cell. Such a late start guarantees smaller discreteness errors \citep[see][for more details]{2021MNRAS.500..663M}. For the present work, we focus on the snapshots at $z = 0, 1,$ and 2.

\subsection{The \ELEPHANT simulation suite}

The Extended LEnsing PHysics using ANalaytic ray Tracing (\ELEPHANT) cosmological simulation suite was run using the \ECOSMOG simulation code \citep{Li:2011vk,Li:2013nua,Barreira:2015xvp,Bose:2016wms}, which is based on the dark matter and hydrodynamic AMR simulation code \RAMSES and includes various types of modified gravity models \citep[e.g.][]{Li:2011vk,2012JCAP...10..002B,Brax:2013mua,Li:2013nua,Li:2013tda,Becker:2020azq}. It is particularly designed to solve for a non-linear scalar field using AMR. New simulations were run for the purpose of testing the effective field theory of large-scale structure (EFTofLSS) pipeline for spectroscopic galaxy clustering \citep{2018MNRAS.476.3195C, 2021JCAP...09..021F,2023arXiv230611053C,KPJC6P4}. For this purpose, 11 simulations were carried out using the \euclid reference cosmology without massive neutrinos for $\Lambda$CDM and the nDGP model (Table 2 of \citealp{2021MNRAS.505.2840E}).
The cosmological parameters of the \lcdm simulations are: $\om = 0.319$, $\ob = 0.049$, $\ol =0.681$, $H_0=67.0\kmsmpc$, $A_{\rm s} = 2.1 \times 10^{-9}$, and $n_{\rm s}=0.96$.
The nDGP simulations used the same parameters as the \lcdm simulations with the cross-over scale $r_{\rm c} = 1.2 \, c / H_0$.
All of the simulations in this simulation suite had a box size of $1024\mpcoh$ and $1024^3$ particles. The initial conditions were generated at $z_{\rm init} = 49$ with 2LPT using the \texttt{FML} code\footnote{\href{https://github.com/HAWinther/FML}{\faicon{github}~https://github.com/HAWinther/FML}} with fixed initial amplitudes. The phases of 10 realisations were extracted with different random seeds, while one realisation shares the same random seed as one of the other simulations, but with opposite phases to have a single paired-and-fixed simulation pair with suppressed cosmic variance \citep{AnguloPontzen16}. Output redshifts were selected from the Euclid Collaboration forecast paper for galaxy clustering \citep[][$z=1.0$, $1.2$, $1.4$ and $1.65$]{Euclid:2019clj}. 

\subsection{The \ColaHiRes simulations}

This simulation series contains overall seven simulations in \lcdm and nDGP cosmologies that were run with \texttt{MG-COLA}, a modified gravity extension of the COmoving Lagrangian Acceleration (COLA) algorithm as implemented in the \texttt{FML} code. The COLA method uses a combination of analytic 2LPT displacement and particle mesh (PM) simulations to perform fast approximate simulations \citep{2013JCAP...06..036T}. These techniques are extended to modified-gravity models using approximate screening methods to preserve the speed advantage of COLA simulations \citep{2017JCAP...08..006W}.
The downside of PM simulations is that the internal structure of dark matter haloes is not well resolved due to limited resolution. This has an important implication for dark matter halo statistics. To mitigate this problem, the COLA simulations were run with an increased mass resolution \citep{Fiorini:2023fjl}.
All simulations in this suite have a box size of $1024\mpcoh$, with $2048^3$ particles. 
The base cosmological parameters of the simulations are the Planck 2015 parameters \citep{2016A&A...594A..13P}: $\om=1-\ol=0.3089$, $\ob=0.0486$, $H_0=67.74\kmsmpc$, $n_{\rm s}=0.9667$, and $\sigma_8=0.8159$. This simulation series focusses on nDGP gravity and tested four cases: $r_{\rm c} =\{0.5,\, 1,\, 2,\, 5\} \, c / H_0$.
The series contains paired-and-fixed simulations \citep{AnguloPontzen16} to suppress cosmic variance in $\Lambda$CDM and in the nDGP model for $r_{\rm c} = 1 \, c / H_0$, while for the others they were only run for a single fixed amplitude realisation.
The initial conditions were generated at $z_{\rm init}=127$ using 2LPT.
Full particle snapshots were stored at four redshift values, $z=1.0$, $1.2$, $1.4$, and $1.65$, motivated by the expected H$\alpha$-emitters redshifts in the \Euclid spectroscopic survey \citep{Euclid:2019clj}.

\subsection{The \DUSTGRAIN and \DUSTGRAINPATHFINDER simulations}

The \DUSTGRAIN (Dark Universe Simulations to Test GRAvity In the presence of Neutrinos) project is an initiative aimed at investigating the degeneracy between $f(R)$ gravity and massive neutrinos at the level of non-linear cosmological observables, which was first pointed out in \citet{Baldi:2013iza}. More specifically, the project includes two suites of cosmological dark-matter-only simulations named the \DUSTGRAIN{\em -pathfinder} \citep[\DUSTGRAINPATHFINDER,][]{Giocoli:2018gqh} and the \DUSTGRAIN{\em -fullscale} simulations that have been run by joining the \MGGADGET \citep[][]{2013MNRAS.436..348P} solver for $f(R)$ gravity and the massive neutrinos implementation \citep[][]{Viel_2010} available within the \PGadgetThree code. The former has been described and validated in \citet{Winther:2015wla} and \citet{KPJC6P1}, while the latter has been compared with other methods in \cite{Adamek23}.

The \DUSTGRAINPATHFINDER simulations have been developed to sample the joint $(f_{R0}, m_{\nu})$ parameter space to identify the most degenerate combinations of parameters with respect to some basic LSS statistics. These include the non-linear matter power spectrum, the halo mass function, weak-lensing-convergence power spectrum, various higher-order statistics, cosmic voids, velocity fields \citep[see][]{Peel:2018aly, Peel:2018aei, Merten:2018bgr, contarini_2021, Garcia-Farieta:2019hal, Hagstotz:2018onp, Hagstotz:2019gsv, Boyle:2020bqn}.
This series includes in total 13 simulations in $f(R)+m_\nu$ cosmology, plus an additional suite of 12 standard $\Lambda $CDM simulations for varying one single standard cosmological parameter at a time that have been specifically run for the Higher-Order Weak Lensing Statistics (HOWLS) project \citep[][]{2023A&A...675A.120E}. These simulations have a box size of $750\mpcoh$ per side, used a softening length of $\varepsilon=20\kpcoh$, and include $(2\times )768^{3}$ particles (for the CDM and neutrinos components). The cosmological parameters (for the reference $\Lambda $CDM cosmology with massless neutrinos) have been set to $\om = 1 - \ol = 0.31345$, $\sigma_8 = 0.842$, $H_0 = 67.31\kmsmpc$, $n_{\rm s} = 0.9658$, and the total matter density has been kept constant when varying the neutrino mass.  Full snapshots have been stored at 34 output times between $z=99$ (corresponding to the starting redshift of the simulation) and $z=0$. 

The \DUSTGRAIN{\em -fullscale} simulations include only three runs (a reference $\Lambda $CDM cosmology and two $f(R)$ gravity models with $f_{R0}=-10^{-5}$ and different values of the total neutrino mass, namely $ m_{\nu }=\{0.1,\, 0.16\}$ eV) simulated in a $8\gpcohcube$ volume containing $(2\times )2048^{3}$ particles. In order to allow for a direct comparison with the \DEMNUni simulations described above, and to produce an extension to the latter for $f(R)$ gravity with massive neutrino cosmologies, the \DUSTGRAIN{\em -fullscale} simulations share the same initial conditions with \DEMNUni for each of the values of the neutrino mass. Therefore, the two sets of simulations have the same statistical realisations of the Universe and identical cosmological parameters. Full snapshots have been stored for 73 output times between $z=99$ (i.e. the initial conditions) and $z=0$.

\subsection{The \CIDER simulations}

The Constrained Interacting Dark EneRgy scenario \citep[or \CIDER,][]{Barros_etal_2019} is a particular type of coupled Quintessence models characterised by a background cosmic expansion that is fixed by construction to be identical to a standard $\Lambda $CDM cosmology. As is discussed in Sect.~\ref{sec:iDE}, this implies refraining from choosing a priori any specific functional form for the scalar self-interaction potential and letting the dynamic evolution of the field sample the potential shape required to match the imposed expansion history. 
The main feature of the \CIDER models is that they show a suppressed growth of structures compared to a standard $\Lambda $CDM model with the same expansion history, thereby possibly easing the $\sigma _{8}$ tension without further exacerbating the tension on $H_{0}$. For these reasons, the model has received some attention even though -- at least in its original form -- it may already be quite tightly constrained by CMB observations \citep[][]{Barros:2022bdv}.
The \CIDER simulations have been run with the \CGADGET code (\citealt{Baldi:2008ay}, see also \citealt{KPJC6P1}) that implements all the relevant features of interacting dark energy models, and includes three values of the coupling $\beta = 0.03, 0.05, 0.08$ besides a reference $\Lambda $CDM cosmology corresponding to the case $\beta = 0$. All simulations clearly share the same expansion history, consistent with the following cosmological parameters: $\om = 1-\ol = 0.311$, $\ob = 0.049$, $H_0 = 67.7\kmsmpc$, $n_{\rm s} = 0.9665$, $A_{\rm s} = 1.992 \times 10^{-9}$, corresponding to a value of $\sigma _{8} = 0.788$ at $z=0$ in the reference $\Lambda$CDM model. The simulations follow the evolution of $2\times 1024^3$ particles for the (coupled) dark matter and (uncoupled) baryon components in a cosmological volume of $1\gpcohcube$ with a softening length of $\varepsilon=25\kpcoh$. The baryonic species are treated as a separate family of collisionless particles, in other words, neither hydrodynamic forces nor radiative processes are considered in the simulations, and its inclusion is required in order to consistently represent the effects of the non-universal coupling characterising these models. Therefore, baryonic particles will interact with other massive particles according to standard Newtonian forces, while the interaction between pairs of CDM particles will be governed by an effective gravitational constant $G_{\rm eff}=G_{\rm N}\left[ 1+ (4/3)\beta ^{2}\right]$ \citep[see e.g.][]{Amendola:2003wa,Baldi:2008ay}.
Full snapshots have been stored for 25 output times between $z=99$ and $z=0$.

\subsection{The \DAKAR and \DAKARTWO simulations}

The dark scattering (DS) scenario \citep[][]{Simpson:2010vh} is another particular class of coupled Quintessence models where a non-universal interaction between dark matter particles and a classical scalar field playing the role of dark energy is characterised by a pure momentum exchange between the two species, with no transfer of rest-frame energy \citep[see e.g.][]{Pourtsidou:2013nha, Skordis:2015yra}. In this respect, this interaction resembles a process of elastic scattering of massive particles (i.e. the dark matter) moving in a homogeneous fluid with an equation-of-state parameter, $w$ (i.e. the dark energy field), which can be simulated by introducing a velocity-dependent force acting on dark matter particles that will depend on the evolution of the dark energy equation-of-state parameter, $w$, and on the cross-section, $\sigma$, characterising the interaction strength \citep[][]{Baldi:2014ica}.

The \DAKAR \citep[][]{Baldi:2016zom} and \DAKARTWO simulations have been run with the \CGADGET code and cover various combinations of the shape of $w(z)$, including the CPL parametrisation given by Eq.~(\ref{eq:wzCPL}) and hyperbolic tangent shapes, and of the cross-section, $\sigma $, giving rise to a diverse phenomenology at both linear and non-linear scales. In particular, DS models have been shown to suppress the linear growth of perturbations for equation-of-state parameters $w>-1$ \citep[][]{Pourtsidou:2016ico, Bose:2017jjx, Carrilho:2021rqo}, thereby possibly addressing the $\sigma _{8}$ tension, but such suppression is typically paired with a substantial enhancement of structure growth at deeply non-linear scales.

The \DAKAR simulations are subject to the approximation of considering the entirety of matter in the Universe is in the form of dark matter, thereby slightly overestimating the effect of the interaction as well as not capturing the segregation effects between dark matter and baryons due to the non-universality of the coupling. These have been run for a cosmology with $\om = 1-\ol = 0.308$, $H_0 = 67.8\kmsmpc$, $n_{\rm s} = 0.966$, $A_{\rm s} = 2.215\times 10^{-9}$, in a simulation box with a volume of $1\gpcohcube$ filled with $1024^{3}$ dark matter particles and using a softening length of $\varepsilon=12\kpcoh$.

The \DAKARTWO simulations, instead, share the same cosmology and the same statistical realisation as the \CIDER simulations described above (i.e. the two sets of simulations share exactly the same reference $\Lambda $CDM run) and include collisionless baryons as a separate family of uncoupled particles, thereby consistently capturing the non-universality of the DS interaction. As for the \CIDER simulations, a collection of 25 full snapshots for redshifts between $z=99$ and $z=0$ has been stored.

\subsection{The \ClusteringDE simulations}

The \textsc{Clustering Dark Energy} simulations are run using the \KEVOLUTION code, a relativistic {\it N}-body code \citep{Hassani_2019, Hassani:2019wed} based on \GEVOLUTIONONETWO \citep{Adamek_2016}. In \KEVOLUTION, the field equations for $k$-essence type theories (Eq.~\ref{eq:action_kessence}) are solved using the effective field theory (EFT) framework. We have two free parameters in the EFT framework of these theories: the equation-of-state parameter, $w(\tau)$, appearing at the background level and kineticity, $\alpha_{\rm K}(\tau)$, at the perturbation level. In the fluid picture of these theories, the relevant parameters are the speed of sound, $c_{\rm s}(\tau)$, and the equation-of-state parameter, $w(\tau)$, which in general are time-dependent. The term `clustering dark energy' refers to the fact that these theories include a sound-horizon scale, beyond which scalar-field perturbations can grow.
In the analysed simulations, constant $w_0$ and $c_{\rm s}^2$ are used, with cosmological parameters based on the \euclid reference cosmology \citep{2021MNRAS.505.2840E}. The suite contains one $\Lambda$CDM simulation and four clustering dark energy simulations: $(w_0,\;c_{\rm s}^2)=(-0.9, 1\,c^2)$, $(-0.9, 10^{-4}\,c^2)$, $(-0.9, 10^{-7}\,c^2)$, and $(-0.8, 10^{-7}\,c^2)$. In these simulations, the box size was set to $2\gpcoh$ with $N=1200^3$ particles. Moreover, two sets of simulations with different resolutions were considered to study the convergence of the results. In this high-resolution simulation set, the box size was set to $2\gpcoh$ with $N = N_{\rm grid} = 2400^3$.
In this series, the particle snapshots were saved in \GADGETTWO format at five different redshifts $z \in\{2, 1.5, 1, 0.5, 0\}$.

\subsection{The \FORGE and \BRIDGE simulation suites}

The \FORGE simulation suite \citep{2021arXiv210904984A} is a set of 198 dark matter only simulations for $f(R)$ gravity and \lcdm run with the \AREPO cosmological simulation code \citep{2010MNRAS.401..791S, 2020ApJS..248...32W} using its MG module \citep{2019NatAs...3..945A}. The simulations explore the cosmological and $f(R)$ parameter space spanned by $\om$ ($\ol = 1-\om$), $h$, $\sigma_8$, and $\fRz$ through 50 combinations (nodes) of these parameters sampled in a Latin-hypercube. All other cosmological parameters are fixed to a Planck cosmology \citep[$n_{\rm s} = 0.9652$, $\ob = 0.049199$,][]{2020AA...641A...6P}. For each node, \FORGE consists of a pair of large box simulations with $512^3$ particles in a $1.5\gpcoh$ side-length box and a pair of high-resolution runs with $1024^3$ particles in a $500 \mpcoh$ box. For each pair, the initial conditions were chosen such that the large-scale variance in the 3D matter power spectrum approximately cancels when averaged over the two simulations \citep[see][for further details and some applications of these simulations]{2021arXiv210904984A,Ruan:2023mgq,Harnois-Deraps:2022bie}. All simulations in this suite started from $z_{\rm init} = 127$, with initial conditions generated using the \TWOLPTIC \citep{2006MNRAS.373..369C} code.


The \BRIDGE simulation suite \citep{Harnois-Deraps:2022bie} uses the same base setup and ICs as \FORGE, but for the nDGP gravity model implemented into the \AREPO code \citep{Hernandez-Aguayo:2020kgq}. Accordingly, instead of varying the $\fRz$ parameter, the nDGP parameter, $\horc$ is varied to explore the cosmological parameter space, with all the other parameters identical to those in \FORGE for corresponding nodes.

The fact that the \FORGE and \BRIDGE simulations have the same cosmological parameters, node by node, allows a third suite of simulations to be done as a control set to quantify the effect of modified gravity and how it correlates to the effect of varying cosmological parameters. This additional suite of simulations, \FORGE-\lcdm, uses the same setup but runs for the \lcdm{} counterparts of the corresponding $f(R)$ and nDGP models. 

As the \FORGE, \FORGE-\lcdm{}, and \BRIDGE simulations have different cosmologies, the mass resolution differs amongst them. In Table \ref{tab:sims}, we thus quoted an order of magnitude for the mass resolution.


\subsection{The \PNGUNITsim simulation}

The \PNGUNITsim \citep{Adame24} is a twin of one of the existing UNITsims \cite[Universe {\it N}-body simulations for the Investigation of Theoretical models,][]{Chuang19}, but with local primordial non-Gaussianities given by  $f_{\rm NL}=100$.
The simulation assumes the following $\Lambda$CDM parameters: $\om = 1 - \ol = 0.3089, \hspace{0.25cm} H_0 = 67.74\kmsmpc, \hspace{0.25cm} n_{\rm s} = 0.9667, \hspace{0.25cm} \sigma_{8} = 0.8147$. It consists of $N=4096^3$ particles in $L=1 \gpcoh$ evolved with \LGADGET, which is a version of \GADGETTWO optimised for massive parallelisation, using a tree-PM algorithm with a softening length of $\varepsilon=6\kpcoh$.
The initial conditions are run with the 2LPT implementation in the \FASTPM \citep{2016MNRAS.463.2273F} code at $z=99$. Both the \UNITsim and \PNGUNITsim are run with fixed initial conditions \citep{AnguloPontzen16}, which set the amplitude of the ICs to their expected value. Whereas there are four \UNITsim simulations in two sets of pairs (within each pair, each simulation has the inverted phases one with respect to another, following \citealt{AnguloPontzen16}), we only have one simulation for the \PNGUNITsim. The PNG-UNITsim is run with the phases of the ICs matched to one of the UNITsims, which is labelled in the databases as `Ampl1'. The usage of fixed ICs with local PNG was validated in \citet{Avila23}, where it was also shown how to increment the precision of the statistics measured from matched simulations. Overall, 129 snapshots were stored during the simulation, and 32 in the $0.5<z<2.0$ range.

\section{Analysis}
\label{sec:analysis}

\begin{figure}
    \centering
    \includegraphics[width=0.49\textwidth]{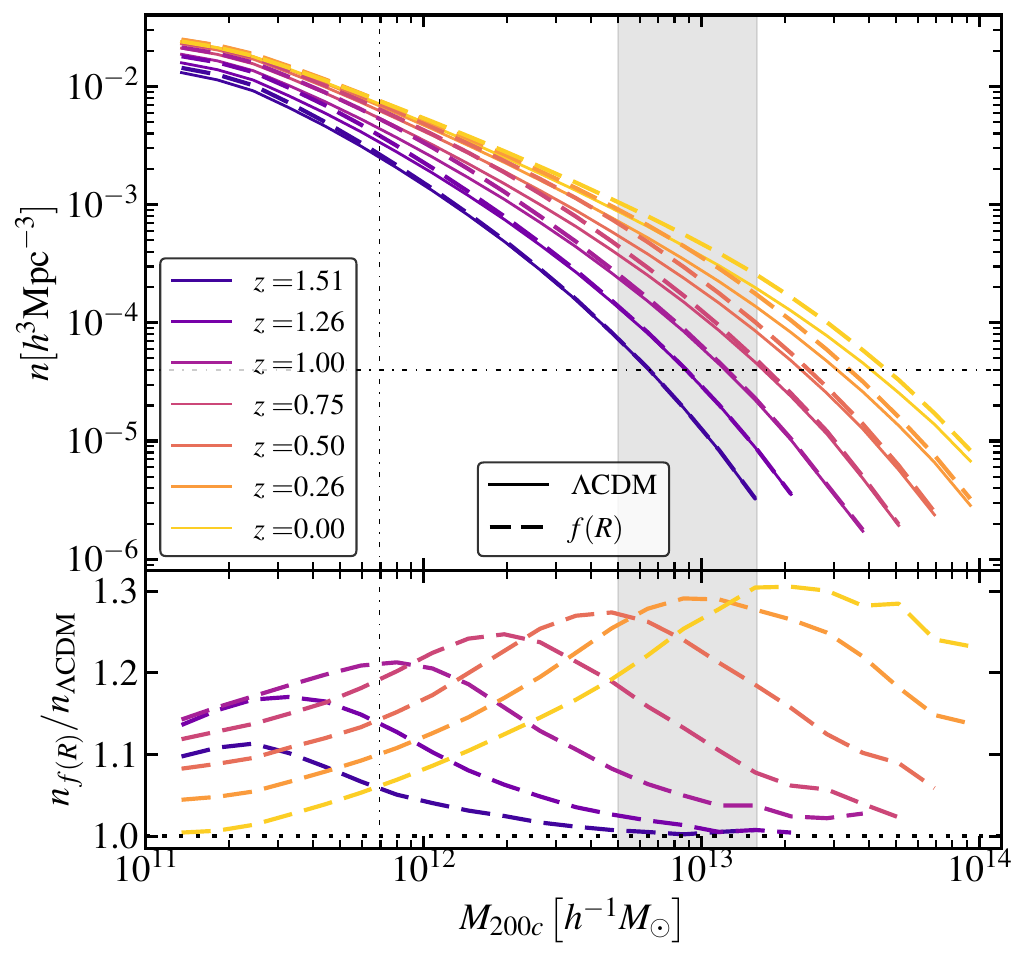}
    \caption{Calculated halo mass function from the \FORGE simulation suite. The vertical and horizontal dash-dotted lines indicate the mass relative to halos with 50 particles and the number density relative to a $1\%$ shot-noise error, respectively. The shaded region highlights the mass bin used to calculate the halo power spectra shown in Fig.~\ref{fig:ForgePkComparison}.} 
    \label{fig:ForgeHMFComparison}
\end{figure}

\begin{figure*}
    \centering
    \includegraphics[width=0.99\textwidth]{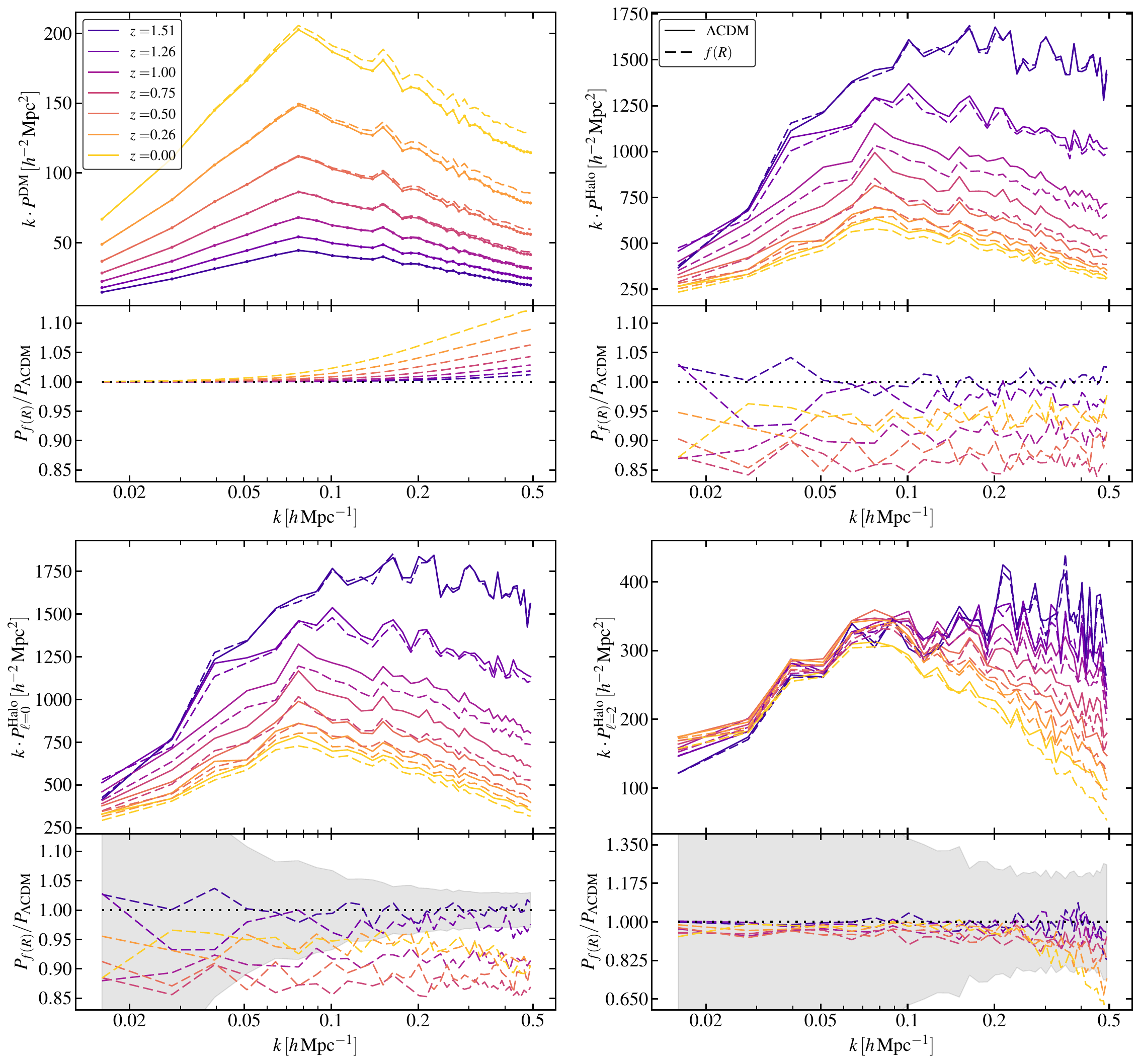}
    \caption{Calculated matter and halo power spectra from the \FORGE simulation suite in the mass bin $10^{12.7}\msoh<M_{\rm halo}<10^{13.2}\msoh$. The solid lines represent the reference \lcdm simulations, while the dashed lines are the results of the nDGP simulations. \textit{Top left:} Real-space power spectra for dark matter. The dots above the solid lines highlight the locations where the power spectrum is estimated. \textit{Top right:} Real-space power spectra for haloes. \textit{Bottom left:} Monopole of the halo power spectrum in redshift space. The shaded region shows the calculated variance of the monopole of the \lcdm halo power spectrum at redshift $z=0$. \textit{Bottom right:} Quadrupole of the halo power spectrum in redshift space. The shaded region represents the calculated variance of the quadrupole of the \lcdm halo power spectrum at redshift $z=0$.} 
    \label{fig:ForgePkComparison}
\end{figure*}

\begin{figure}
    \centering
    \includegraphics[width=0.49\textwidth]{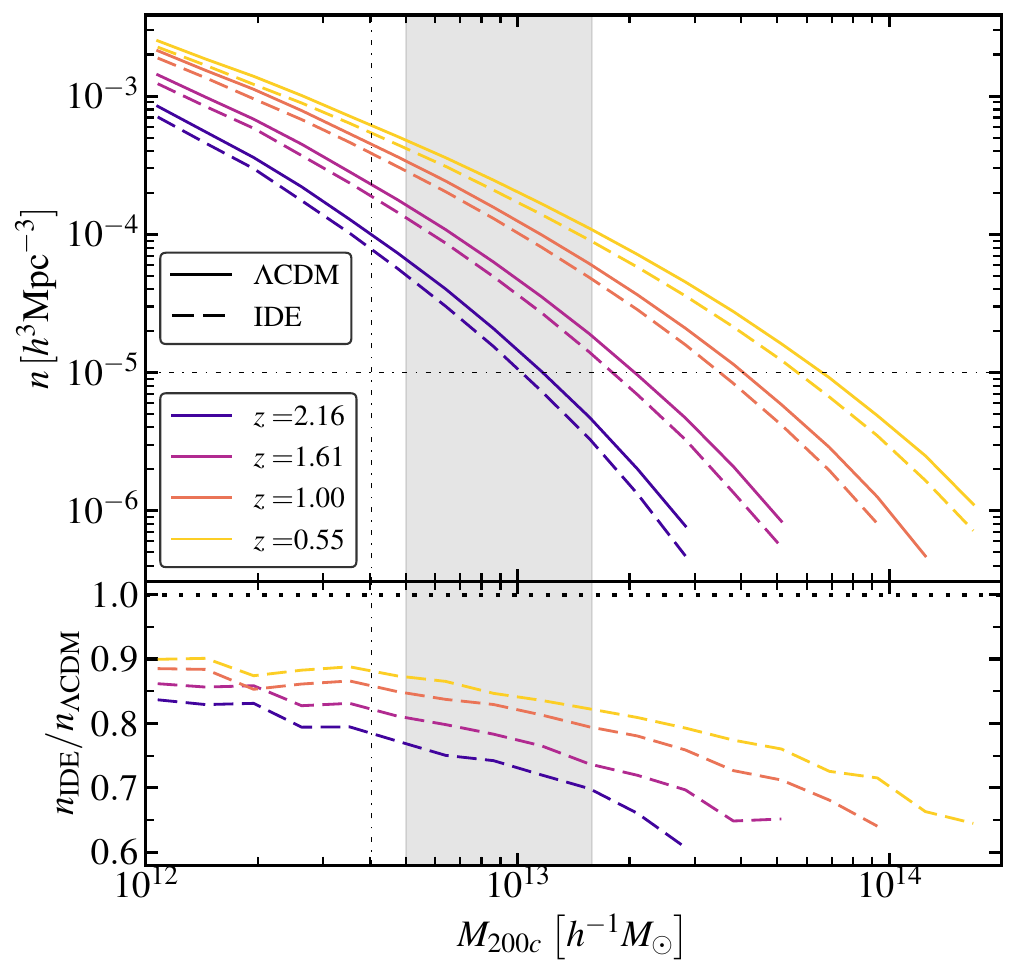}
    \caption{Calculated halo mass function from the \CIDER simulation suite. The vertical and horizontal dash-dotted lines indicate the mass relative to halos with 50 particles and the number density relative to a $1\%$ shot-noise error, respectively. The shaded region highlights the mass bin used to calculate the halo power spectra shown in Fig.~\ref{fig:CiderPkComparison}.} 
    \label{fig:CiderHMFComparison}
\end{figure}

\begin{figure*}
    \centering
    \includegraphics[width=0.99\textwidth]{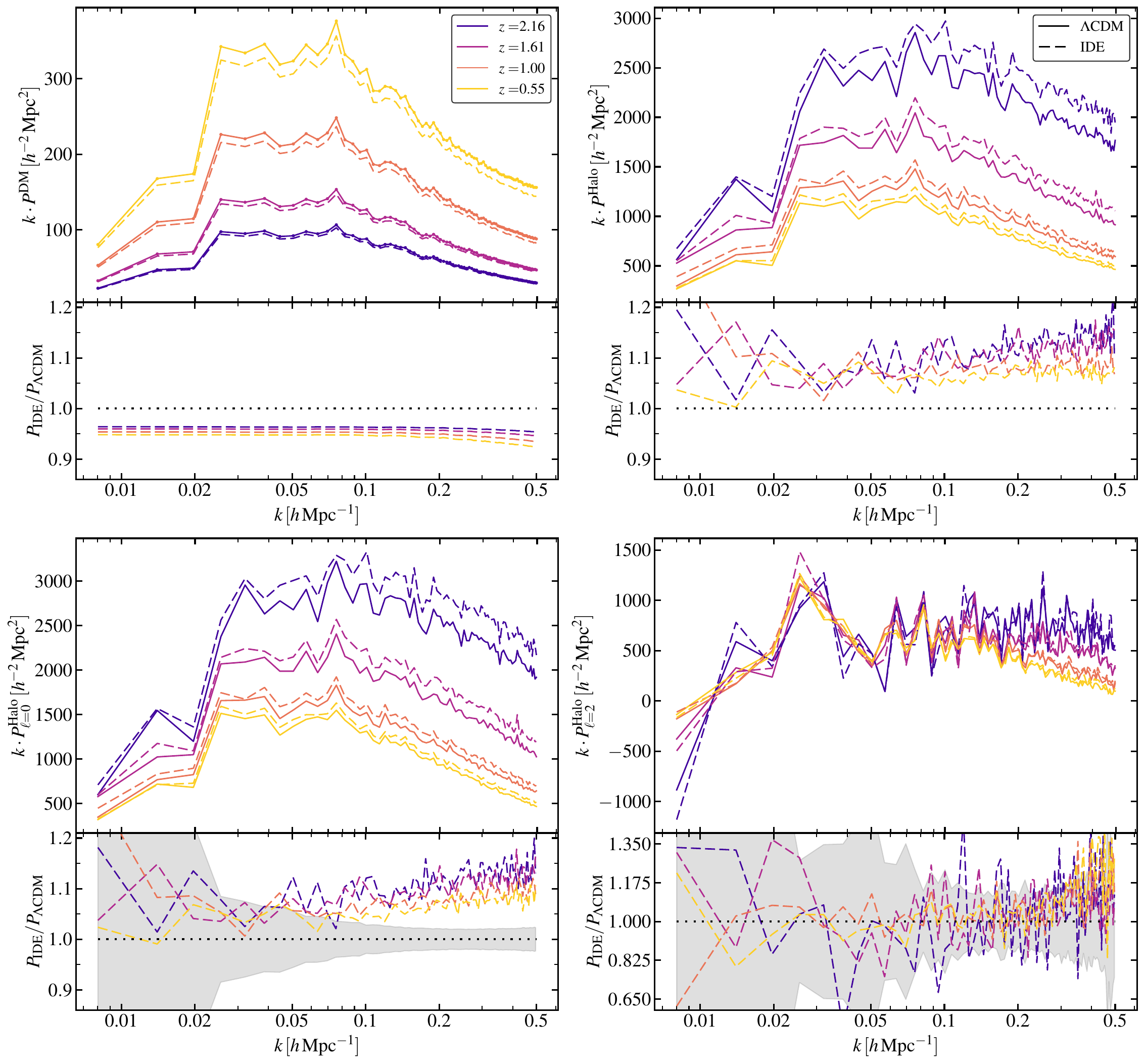}
    \caption{Calculated matter and halo power spectra from the \CIDER simulation suite in the mass bin $10^{12.7}\msoh<M_{\rm halo}<10^{13.2}\msoh$. The solid lines represent the reference \lcdm simulations, while the dashed lines are the results of the IDE simulations. \textit{Top left:} Real-space power spectra for dark matter. The dots above the solid lines highlight the locations where the power spectrum is estimated. \textit{Top right:} Real-space power spectra for haloes. \textit{Bottom left:} Monopole of the halo power spectrum in redshift space. The shaded region shows the calculated variance of the monopole of the redshift space \lcdm halo power spectrum at redshift $z=0.55$. \textit{Bottom right:} Quadrupole of the halo power spectrum in redshift space. The shaded region represents the calculated variance of the quadrupole of the redshift space \lcdm halo power spectrum at redshift $z=0.55$.} 
    \label{fig:CiderPkComparison}
\end{figure*}

We have developed a cosmological analysis pipeline to generate mock observables from non-standard cosmological simulations in a consistent and rapid way. The pipeline is a \texttt{SLURM}\footnote{\url{https://slurm.schedmd.com/}} script that runs in parallel on multiple nodes on the machines where the simulations are stored. The pipeline consists of several modules that can be activated or deactivated independently. The modules are controlled by a configuration file that specifies the input and output parameters, as well as the options for each module. The input of the pipeline is the particle snapshots of the non-standard cosmological simulations. The supported input formats are \GADGET binary and Hierarchical Data Format version 5 \citep[HDF5, ][]{2005MNRAS.364.1105S}, \AREPO HDF5 \citep{2020ApJS..248...32W}, \RAMSES HDF5 format from \RAYGAL \citep{2014A&A...564A..13R,rasera2022raygal}, and \GIZMO HDF5 \citep{2015MNRAS.450...53H}. The main steps of the analysis are summarised in Fig.~\ref{fig:FlowChart}. In this section, we describe the quantities generated by this pipeline.

\subsection{Dark matter density field}
\label{sec:rspacedm}
The pipeline uses \nbodykit \citep{Hand_2018} to read and analyse the dark matter particle data of the input-simulation snapshots. This \texttt{Python} package is an open-source, massively parallel toolkit that provides a set of LSS algorithms useful in the analysis of cosmological data sets from {\it N}-body simulations and observational surveys. During the dark matter density-field analysis, \nbodykit generates a reconstructed density field from the input-particle distribution with the triangular-shaped-cloud (TSC) density-assignment function. We chose to use a
\begin{equation}
    N_{\rm grid} = 2^{\left\{\floor\left[\log_2\left(\sqrt[3]{N_{\rm part}}\right)\right]-1\right\}} 
    \label{eq:gridsize}
\end{equation}
linear grid size for every analysed snapshot, where $N_{\rm part}$ is the number of stored particles in the snapshot. With this choice, there will always be at least eight particles on average in each cubic density cell. The reconstructed density fields were saved in \textit{bigfile} format \citep{yu_feng_2017_1051252} for future analysis.

\subsubsection{Real-space power spectrum}
\label{sec:rspacedmpk}
The real-space matter power spectrum is defined via
\begin{equation}
    \left <\tilde{\delta}(\vec{k})\tilde{\delta}^*(\vec{k}’) \right> = (2\pi)^3\,P(k)\,\delta_{\rm D}^{(3)}(\vec{k} - \vec{k}’)\,,
    \label{eq:rspacepk}
\end{equation}
where $\tilde{\delta}(\vec{k})$ is the Fourier-transform of the matter overdensity field \begin{eqnarray}
    \delta(\vec{r})\,=\,\frac{\rho(\vec{r})}{\overline{\rho}} - 1 \,,
\end{eqnarray} and $\vec{k}$ is the wavevector. We estimated the power spectrum using \nbodykit. The density field was created by binning the particles into a grid using a TSC-density-assignment function, with the linear-grid size defined in Eq.~(\ref{eq:gridsize}). The density field was Fourier-transformed and the power spectrum was computed by binning $|\tilde{\delta}({\vec k})|^2$, deconvolving the window function and subtracting shot noise. We also used the interlacing technique to reduce aliasing \citep{2016MNRAS.460.3624S}. The bin size of the power spectrum was set to
\begin{equation}
    \Delta k = k_{\rm f} = \frac{2\pi}{L_{\rm box}},
    \label{eq:dk}
\end{equation}
where $k_{\rm f}$ is the fundamental wavenumber, and $L_{\rm box}$ is the linear size of the simulation. The pipeline saves the power spectrum of every calculated bin below the
\begin{equation}
    k_{\rm Ny} = \frac{\pi N_{\rm grid}}{L_{\rm box}}
\end{equation} 
Nyquist wavenumber with the number of modes into a simple ASCII format file.

\subsubsection{Redshift-space power spectrum}
\label{sec:zspacedmpk}
The real-space matter power spectrum is not directly measurable in galaxy surveys because we cannot probe the real-space positions of galaxies. What we can directly measure is the redshift-space power spectrum $P^s(k,\mu)$ where $\mu = \hat{\vec{n}}_{\rm LOS} \cdot \hat{\vec{k}}$  and $\hat{\vec{n}}_{\rm LOS}$ is a unit vector in the line-of-sight (LOS) direction. This can be expanded in multipoles, $P^s(k,\mu) = \sum_{\ell=0}^\infty P_\ell^s(k)\mathcal{L}_\ell(\mu)$, where $\mathcal{L}_\ell(\mu)$ are the Legendre polynomials. The multipoles are then computed from the redshift-space power spectrum as
\begin{equation}
P_\ell^s(k) = \frac{2\ell +1}{2}\int_{-1}^1 P^s(k,\mu) \mathcal{L}_\ell(\mu)\,{\rm d}\mu.
    \label{eq:zspacepkmultipoles}
\end{equation}
We computed the redshift-space power spectrum in 25 $\mu$ bins and the redshift-space multipoles (the monopole $P_0$, the quadrupole $P_2$, and the hexadecapole $P_4$) using \nbodykit from the input dark matter density field. For this, we used the distant-observer approximation
\begin{equation}
    \vec{s}_{i} = \vec{r}_{i} + \left( \hat{\vec{n}}_{\rm LOS} \cdot \frac{\vec{v}_i}{a H} \right) \cdot \hat{\vec{n}}_{\rm LOS},
    \label{eq:rsd}
\end{equation}
to add the redshift-space distortions using the three coordinate axes as the LOS directions (observables were computed as the mean over these three individual axes). Here, $\vec{r}_i$ and $\vec{v}_i$ are the real-space particle coordinates and peculiar velocities inside the periodic simulation box, $\vec{s}_i$ is the corresponding redshift-space position we computed, $a=1/(z+1)$ is the scale factor, and $H$ is the Hubble parameter at the redshift of the snapshot. We deconvolved the window function for the density assignment, applying interlacing, and the resulting density power spectrum was finally shot-noise subtracted. The saved wavenumber bins are the same as in Sect.~\ref{sec:rspacedmpk}.

\subsubsection{Linear dark matter power spectrum}
\label{sec:lindmpk}
During the analysis of scale-independent models, multiple modules use the linear dark matter power spectrum as an input. To make the analysis more transparent, we also generated and saved the linear power spectrum for all analysed redshifts. For this, the pipeline needs the $P_{\rm lin}(k,z_{\rm start})$ input linear power spectrum of the simulation that was used during the initial condition generation. This can be defined at any $z_{\rm start}$ redshift. Then, this linear power spectrum is renormalised with the cosmological parameter $\sigma_8$ at $z=0$. The normalised $P_{\rm lin}(k,z=0)$ power spectrum is rescaled to $z_{\rm snap}$ redshifts of all analysed snapshots as
\begin{equation}
    P_{\rm lin}(k,z_{\rm snap}) = \left[\frac{D(z_{\rm snap})}{D(z=0)}\right]^2 P_{\rm lin}(k,z=0) \;,
    \label{eq:linpk}
\end{equation}
where $D(z)$ is the linear growth function. The pipeline uses this back-scaling since the linear growth in these Newtonian simulations follows this scale-independent evolution. For \lcdm reference simulations, we used the
\begin{equation}
    D(a) = \frac{5\om H_0^2}{2}H(a)\int\limits_{0}^{a}\frac{\diff a'}{\dot{a}'^3}
    \label{eq:LCDMgrowth}
\end{equation}
linear growth to scale the linear spectrum \citep{1993ppc..book.....P}. This growth function only describes the linear growth in the \lcdm framework. In the case of \wcdm and CPL models, we solve the
\begin{equation}
    G'' + \left[\frac{7}{2} + \frac{3}{2}\frac{w(a)}{1+X(a)} \right]\frac{G'}{a} + \frac{3}{2} \frac{1-w(a)}{1+X(a)}\frac{G}{a^2} = 0
    \label{eq:wCDMgrowth}
\end{equation}
ordinary differential equation \citep{2003MNRAS.346..573L} with the \COLOSSUS python package \citep{2018ApJS..239...35D}, where $G(a) = D(a)/a$ and
\begin{equation}
    X(a) = \frac{\om}{1-\om} {\rm e}^{-3\int\limits_{a}^{1}\diff a'\frac{w(a')}{a'}}.
    \label{eq:Xa}
\end{equation}
For every other model, we use tabulated linear growth functions. The linear power spectra are calculated in the same wavenumber bins as the non-linear real-space matter power spectra.

\subsection{Halo catalogues}
\label{sec:halo_catalogue}

\rockstar \citep{Behroozi_2012} is a friends-of-friends (FoF) halo-finder algorithm that uses information from the full 6D phase space (positions and the velocities) of the particles. The code initially creates FoF groups in real space, with a large linking length ($b \simeq 0.28$). It then does a new FoF search using the phase-space metric 
\begin{equation}
    d = \sqrt{\frac{|{\bf x}_1 - {\bf x}_2|^2}{\sigma_x^2} + \frac{|{\bf v}_1 - {\bf v}_2|^2}{\sigma_v^2}} \,,
    \label{eq:drockstar}
\end{equation}
where $\sigma_x$ and $\sigma_v$ are the particle-position and velocity dispersions for the given FoF group. Finally, it links particles into subgroups and this is done iteratively on each subgroup creating a hierarchical set of structures. By default, the algorithm calculates halo and subhalo masses using dark matter particles from the spherical regions around the friends-of-friends group with gravitationally unbound particles removed. The halo masses calculated this way are called bound-only (BO) masses. If the unbound particles are not removed during the mass calculation, the calculated masses are strict spherical-overdensity (SO) masses.

We made a custom version of the publicly available \rockstar code \citep{2012ascl.soft10008B} to analyse our non-standard simulations. We added new input formats for simulations, options to read tabulated expansion histories, and already internally computed quantities in the outputs such as halo minor axis vectors and radii at different mass definitions. None of these modifications impact the halo-finding algorithm.
In our pipeline, we use the following mass-definitions: $M_{\rm 200c}$ (SO \& BO), $M_{\rm 500c}$ (SO), $M_{\rm 1000c}$ (SO), $M_{\rm 2500c}$ (SO), $M_{\rm 200b}$ (SO). 
The $M_{\rm vir}$ masses are not calculated by the pipeline, since this mass definition is dependent on the cosmological parameters and on the laws of gravity. Many non-standard cosmological models are changing the dynamics of the dark matter component, and this choice simplifies the future expansion of the database without the need of implementing new cosmologies in \rockstar. After the catalogue (in ASCII format) is produced, we run a post-processing script to find parent haloes for subhaloes and store the information as an additional index column. Extra information is saved in the header such as scale factor, box length, and particle mass. Additional particle data for each halo are also saved by \rockstar in a custom BGC2 binary data format. During the execution of our pipeline, these BGC2 files are temporarily stored to provide additional input for other analysis modules.

\subsubsection{Halo mass function and power spectra}
\label{sec:rspacehalopk}
By default, we compute the halo mass function (HMF) in the range $11 < \logten[M/(\msoh)] < 14$ with 32 logarithmic bins using the main mass definition (200c) and excluding substructures. The pipeline allows the user to use different mass definitions (see Sect.~\ref{sec:halo_catalogue}), include substructures, and vary the HMF range and binning.

In practice, many simulations produce very large ASCII files that are not practical to read using standard libraries such as \numpy\footnote{\url{https://numpy.org/}} or \pandas.\footnote{\url{https://pandas.pydata.org/}} To speed up the analysis pipeline, we therefore use \polars,\footnote{\url{https://pola.rs/}} a fast multi-threaded dataframe library.

The halo real-space and redshift-space power spectra are computed using the same tools as in Sects. \ref{sec:rspacedmpk} and \ref{sec:zspacedmpk}. The user can specify the halo mass range, SO or BO for the main mass definition, and whether or not to include substructures.

\subsubsection{Halo bias}
\label{sec:halo_bias}
For each catalogue, we infer the linear halo bias with the estimator
\begin{equation}
    b = \left\langle\sqrt{\frac{P_{\rm h}(k)}{P_{\rm m}(k)}}\,\right\rangle_{k<k_{\rm max}},
\end{equation}
with $P_{\rm m}(k)$ and $P_{\rm h}(k)$ the matter and halo real-space power spectra, respectively estimated in Sects. \ref{sec:rspacedmpk} and \ref{sec:rspacehalopk}. This estimator calculates the bias by taking the square root of the ratio of the halo power spectrum $P_{\rm h}(k)$ to the matter power spectrum $P_{\rm m}(k)$, and then averaging this ratio over all $k$ bins where $k < k_{\rm max}$ with uniform weighting. We only use this computed quantity to be as model-independent as possible and to remove cosmic variance (since matter and halo both share the same sample and cosmic variance). We compute the mean power spectra ratio up to a conservative value of $k_{\rm max} = 0.1\homopc$ to mitigate the effects of non-linear clustering. This method works reliably for scale-independent bias with sub-percent accuracy, but cannot be used for models that have scale-dependent bias at $k<k_{\rm max}$ wavenumber.

\subsubsection{Redshift-space Gaussian covariance}

To produce Gaussian covariances of the power-spectrum multipoles in redshift space, we need the linear bias and power-spectrum multipoles including the shot-noise contributions as inputs. The former is estimated numerically in Sect.~\ref{sec:halo_bias}, while the latter can be internally computed from an input linear power spectrum; in this case, the covariance is estimated as in \cite{taruya2010baryon}, or with the EFT model using the \COMET emulator \citep{eggemeier2023comet} and the covariance formulae from \citet{grieb2016gaussian}.

When analysing snapshots, it may be interesting to compute the power-spectrum multipoles averaged over the three box directions to significantly suppress variance. However, this procedure also has to be carefully accounted for in the covariance \citep{smith2021reducing} since the LOS-averaged covariance is not equal to the single-LOS covariance divided by three (as one might naively expect). Thanks to the LOS-averaged covariance implemented in \COMET, it can also be part of the outputs of our pipeline.

\subsubsection{2D and 3D halo profiles}
The generation of binned three-dimensional and two-dimensional projected profiles was performed using a custom analysis module that reads the halo catalogues as well as the BGC2 particle data, and stores the resulting profiles of each halo in a separated HDF5 file.

All the profiles are obtained considering 50 log-spaced bins in a fixed radial range $[0.001,5]$,  in units of $r_{\rm 500c}$. This analysis module provides cumulative mass, density, number density, and cumulative number density profiles, as well as velocity and velocity dispersion profiles for the cartesian- and spherical-coordinates components of the velocities.
The 2D profiles correspond to projecting the LOS along each of the cartesian coordinates in a cylinder of length $5r_{\rm 500c}$. In an upcoming version of the pipeline, the profiles will also be available for the projections along the axes of the inertia ellipsoid $a$, $b$, and $c$.

\subsection{Cosmic voids}

The Void IDentification and Examination toolkit, \href{https://bitbucket.org/cosmicvoids/vide_public/src/master/}{\VIDE} \citep{sutter2014vide}, is a parameter-free topological void finder, conceived for galaxy-redshift surveys and {\it N}-body simulations. The \VIDE pipeline is an open-source \texttt{Python/C++} code, based on the \href{http://skysrv.pha.jhu.edu/~neyrinck/voboz/}{\ZOBOV} \citep{Neyrinck_2008} software and can be launched on any tracer distribution. The algorithm follows the following main steps: i) estimation of the density field of a tracer distribution using the Voronoi tessellation~\citep{2000A&A...363L..29S}; ii) detection of all the relative minima; iii) merging of nearby Voronoi cells into zones via the watershed transform~\citep{2007MNRAS.380..551P}, cells correspond to local catchment `basins', which are identified as voids. \VIDE can also merge adjacent voids to construct a nested hierarchy of voids if a merging threshold is provided. In this case, when two adjacent voids have at least one Voronoi cell on the ridge separating them lower than the threshold, they are merged into a parent void. In this work, in order to leave the algorithm parameter-free, and for consistency with other Euclid void analyses \citep{Hamaus22,Contarini22}, we do not explore this possibility. 

\VIDE provides some fundamental properties of voids. The void size is measured by the effective radius, defined as the radius of a sphere with the same volume as the void, $R_{\rm eff} = [ (3/4 \pi)  \sum_i V_i ]^{1/3}$, where $V_i$ is the volume of the $i^{\rm th}$ Voronoi cell belonging to the void. The void centre is defined as the volume-weighted barycentre, $\Vec{X}_v = \sum_i \vec{x}_i V_i/V_{\rm tot}$, with $V_{\rm tot}=\sum_i V_i$. We note that this corresponds to the geometric centre of the void. In addition, \VIDE also provides the position of the tracer sitting in the lowest-density Voronoi cell, that is, the minimum. The void's depth is estimated via the central density, defined as the mean density in a sphere centred in the barycenter, $\Vec{X}_v$, with radius $R_{\rm eff}/4$. \VIDE also computes void shapes via the inertia tensor as well as the corresponding eigenvalues and eigenvectors. The ellipticity is then computed as $\epsilon = 1- (J_1/J_3)^{1/4}$, where $J_1$ and $J_3$ are the smallest and largest eigenvalues.

We detect voids in the distribution of \rockstar haloes. After the void catalogues are produced, we post-process them to measure the void-size function, which is the number density of voids as a function of their size, $R_{\rm eff}$. The void-size function is a sensitive probe for cosmology, strongly complementary to the galaxy 2pt-statistics~\citep{pisani_2015_abundance,massara_2015,Kreisch2019,verza_2019,contarini_2021,Contarini22,contarini_2022_sdss,Verza_2022a,Verza_2022b,verza_2024}. Additionally, albeit not computed for this paper, the void catalogues allow one to compute the void-galaxy cross-correlation function, another powerful statistic to constrain cosmology \citep[see e.g.][]{Hamaus22}.


\section{Interpretation}
\label{sec:interpretation}

The mock observables that we compute contain several interesting signatures of non-standard models. Due to a large number of analysed models, in this paper, we focus on showing results from the nDGP, $f(R)$, and interacting-dark-energy models, which we obtained thanks to the analysis of the \ELEPHANT, \FORGE, and \CIDER simulations suites, respectively.

To mitigate the noise due to sample variance in the figures below, we average the signals over the available realisations of the \ELEPHANT simulations. In the case of the \FORGE suite, we focus on a single cosmology\footnote{In particular we use node 10 of the Latin-Hypercube sampling described in Table 1 of \cite{2021arXiv210904984A}.} corresponding to $\left|\bar{f}_{\mathrm{R}0}\right| = 10^{-5.34219}$. In the case of the \CIDER simulations, we focus on the IDE model with $\beta=0.03$ coupling. The nDGP, $f(R)$, and IDE simulations were run with the same initial conditions as their \lcdm counterpart so the effects of non-standard cosmologies can be studied directly by comparing the generated observables.

The main difference between the modified-gravity simulation and \lcdm is the inclusion of a fifth force. For nDGP, this fifth force acts on all scales and increases in strength toward redshift zero (and goes to zero as we go to higher and higher redshifts). This is also true for $f(R)$, with the exception that the fifth force has only a finite range, so it does not affect the clustering on the largest scales.

In the IDE simulations, the dark energy component interacts with the dark matter, resulting in a transfer of energy between the two. This interaction affects the growth of cosmic structures by modifying the gravitational potential. This cosmological scenario is expected to suppress structure formation at late time compared to the standard \lcdm model.

These differences lead to a number of different observable signatures, a few of which we highlight below.

{\bf Abundance of dark matter haloes} -- In Fig.~\ref{fig:ElephantHMFComparison}, we compare the cumulative halo mass function of the nDGP and \lcdm models. The inclusion of the fifth force means structures will form more rapidly than in \lcdm and this is indeed what we see. This is most pronounced at the high-mass end where the abundance is up to $50$\% larger. 
In Fig.~\ref{fig:ForgeHMFComparison}, we compare the cumulative halo mass function of the $f(R)$ and \lcdm model. We see roughly the same qualitative features as for nDGP in that the halo abundance generally increases with halo mass and with redshift compared to the reference \lcdm results. However, as opposed to nDGP, we see an over-abundance of `small' haloes at earlier times in the $f(R)$ simulations. This is a consequence of the fact that the fifth force only acts on `small' scales and the fact that the screening mechanism is more effective at suppressing the fifth force in and around the most massive haloes.
The comparison of the halo mass function between the standard \lcdm and the $\beta = 0.03$ IDE model can be seen in Fig.~\ref{fig:CiderHMFComparison}. In the IDE model, the interaction between the dark energy and dark matter caused a substantial reduction in the HMF. This is a straightforward consequence of the suppressed growth rate of the matter fluctuations.

{\bf Clustering of dark matter} -- In Figs.~\ref{fig:ElephantPkComparison}, \ref{fig:ForgePkComparison}, and~\ref{fig:CiderPkComparison}, we show the calculated real- and redshift-space power-spectrum multipoles for the haloes and dark matter for nDGP, $f(R)$, and IDE, respectively. For the modified gravity models, the effect of the fifth force is again clearly in the dark matter power spectrum and shows two different effects: for nDGP, we have a scale-independent growth rate causing the power spectrum to be boosted on all scales displayed, while for $f(R)$ we have a scale-dependent growth-rate where $f(R)$ agrees with \lcdm on the largest scales, but is boosted below a critical scale that is related to the range of the fifth force. For both models, the difference with respect to \lcdm increases in strength as we get closer to the present time. 
In the case of the interacting-dark-energy model, the energy transfer between the dark energy and dark matter caused a scale-independent suppression in the dark matter power spectrum. At redshift $2$, this is $\simeq3$\%, and this difference increases to $\simeq5$\% for $z=0.55$ compared to the \lcdm model.

{\bf Halo bias} -- When it comes to halo clustering in real space, we see the opposite effect as for the dark matter power spectrum in Figs.~\ref{fig:ElephantPkComparison} and~\ref{fig:ForgePkComparison}, with nDGP and $f(R)$ being less clustered than \lcdm. This comes from a smaller halo bias in these modified-gravity models (see e.g. \citealt{Barreira_2014} for a theoretical explanation for nDGP).
In the interacting-dark-energy scenario, the halo bias in real space is $6\%$ higher than for \lcdm. As a consequence, the real-space clustering of the dark matter haloes is more prominent in the IDE simulation.

{\bf Redshift space distortions} -- For the redshift-space halo power spectra, the boost in the ratio with respect to \lcdm is seen to be larger than in real space, which comes from the larger velocities in the modified-gravity simulations, leading to enhanced redshift-space distortions. The monopole redshift-space power spectra of the haloes in the mass bin $10^{12.7}\msoh<M_{\rm halo}<10^{13.2}\msoh$ in the IDE simulations are showing a $5-10$\% excess power compared to the \lcdm counterpart, similarly to the real-space clustering. On the other hand, the quadrupole only shows a power increase at smaller, non-linear scales.

\section{Summary}
\label{sec:summary}
In this paper, we describe a new pipeline based on the \rockstar halo finder and the \nbodykit LSS toolkit to post-process cosmological simulations with modified gravity, non-standard expansion history, modified dark matter or dark energy components, or altered initial conditions. We used this pipeline to analyse 474 cosmological {\it N}-body simulations in various \lcdm and non-standard cosmological scenarios in a consistent way. With this pipeline, we generated halo catalogues, halo mass functions, reconstructed density fields, real- and redshift-space power spectra, Gaussian covariances, halo biases, and void catalogues. This generated data will serve as a theoretical prediction and reference for \euclid as well as other Stage-IV cosmology projects. Using the calculated quantities, we identified distinctive signatures of non-standard behaviour in nDGP and $f(R)$ modified-gravity models, and in the \CIDER interacting-dark-energy scenario.
 
The synthetic halo catalogues are crucial in the production of additional observables, which can be used for a direct comparison with cosmological observations of \euclid.
In the near future, we shall extend the generated database with halo density profiles \citep{1996ApJ...462..563N, 2018MNRAS.473L..69L}, synthetic galaxy catalogues \citep{2003ApJ...593....1B}, weak lensing \citep{1990ApJ...365...22J, 2001PhR...340..291B} and ISW \citep{2008PhRvD..77l3520G} maps, and lightcones \citep{2013MNRAS.429..556M}.

 We have generated overall more than $100$ TB of post-processed data from the available non-standard simulations. During the analysis, the pipeline used 66 CPU hours and 60GB of memory per billion particles per snapshot on average. The data are available on request on the CosmoHub \footnote{\url{https://cosmohub.pic.es/home}} platform \citep{2020A&C....3200391T, 2017ehep.confE.488C} designed for interactive exploration and distribution of massive cosmological datasets.

\begin{acknowledgements}
\label{sec:acknowledgements}
\AckEC
GR’s research was supported by an appointment to the NASA Postdoctoral Program administered by Oak Ridge Associated Universities under contract with NASA. GR and AK were supported by JPL, which is run under contract by the California Institute of Technology for NASA (80NM0018D0004). GR acknowledges the support of the Research Council of Finland grant 354905.
The authors acknowledge the Texas Advanced Computing Center (TACC) at The University of Texas at Austin for providing HPC and visualization resources that have contributed to the research results reported within this paper. URL: \url{http://www.tacc.utexas.edu}.
This project was provided with computer and storage resources by GENCI at TGCC thanks to the grant 2023-A0150402287 on Joliot Curie's SKL partition.
DFM thanks the Research Council of Norway for their support and the resources provided by UNINETT Sigma2 -- the National Infrastructure for High-Performance Computing and Data Storage in Norway. 
\AckCosmoHub
ZS acknowledges funding from DFG project 456622116 and support from the IRAP and IN2P3 Lyon computing centers. 
During part of this work, AMCLB was supported by a fellowship of PSL University-Paris Observatory.
CG thanks the support from INAF theory Grant 2022: Illuminating Dark Matter using Weak Lensing by Cluster Satellites, PI: Carlo Giocoli.
VGP is supported by the Atracci\'{o}n de Talento Contract no. 2019-T1/TIC-12702 granted by the Comunidad de Madrid in Spain. VGP, and by the Ministerio de Ciencia e Innovaci\'{o}n (MICINN) under research grant PID2021-122603NB-C21.
PNG-UNITsim was run thanks to the MareNostrum supercomputer in Spain and the Red Española de Supercomputación through grants: RES-AECT-2021-3-0004, RES-AECT-1-0007 \& RES-AECT-2022-3-0030.
We extend our sincere gratitude to Christian Arnold and Claudio Llinares for their valuable contributions to this research. Their work significantly influenced the development of this project.
\end{acknowledgements}

\bibliography{Euclid} 

\end{document}